\documentclass[structabstract]{aa}

\usepackage{natbib}
\bibpunct{(}{)}{;}{a}{}{,}
\usepackage{graphicx}
\usepackage{txfonts}
\usepackage{lscape}
\usepackage{longtable}
\usepackage{multirow}
\usepackage[pdftex,pdfpagemode={UseOutlines},bookmarks,bookmarksopen, colorlinks,linkcolor={blue},citecolor={blue},urlcolor={red}]{hyperref}

\begin{document}

\title{A Wide-Field Near-Ir H$_2$ 2.122$\mu$m line survey of the Braid Nebula Star Formation Region in Cygnus~OB7}

\author{%
T.~Khanzadyan\inst{\ref{mpifr}}\and %
C.~J.~Davis\inst{\ref{jach},\ref{nasa}}\and %
C.~Aspin\inst{\ref{unihawaii}}\and %
D.~Froebrich\inst{\ref{kent}}\and %
M.~D.~Smith\inst{\ref{kent}}\and %
T.~Yu.~Magakian\inst{\ref{bao}}\and %
T.~Movsessian\inst{\ref{bao}}\and %
G.~H.~Moriarty-Schieven\inst{\ref{herzberg}}\and %
E.~H.~Nikogossian\inst{\ref{bao}}\and %
T.-S.~Pyo\inst{\ref{subaru}}\and %
T.~L.~Beck\inst{\ref{stsci}}}

\institute{
Max-Plank-Institut f\"{u}r Radioastronomie, Auf dem H\"{u}gel 69, D-53121 Bonn, Germany\\ \email{tkhanzadyan@mpifr-bonn.mpg.de} \label{mpifr}
\and
Joint Astronomy Centre, 660 N. A`oh\=ok\=u Place,  University Park, Hilo, HI 96720, USA \label{jach}
\and
Astrophysics Division, NASA HQ, 300 E Street SW, Mail Stop 3Y28 Washington DC 20546, USA \label{nasa}
\and
Institute for Astronomy, University of Hawaii, 640 N. A`oh\=ok\=u Place, Hilo, HI 96720, USA \label{unihawaii}
\and
Centre for Astrophysics \& Planetary Science, School of Physical Sciences, The University of Kent, Canterbury CT2 7NH, UK \label{kent}
\and
V. A. Ambartsumyan Byurakan Astrophysical Observatory, 0213 Aragatsotn reg., Armenia \label{bao}
\and
NRC Herzberg Institute of Astrophysics 5071 West Saanich Road Victoria, British Columbia, V9E 2E7, Canada \label{herzberg}
\and
Subaru Telescope, National Astronomical Observatory of Japan, 650 N. A`oh\=ok\=u Place, Hilo, HI 96720, USA \label{subaru}
\and
Space Telescope Science Institute, 3700 San Martin Drive, Baltimore, MD 21218, USA \label{stsci}
}

\date{Received ; accepted }


\abstract
	{Outflows and jets are the first signposts of ongoing star formation processes in any molecular cloud, yet their study in optical bands provides limited results due to the large extinction present. Near-infrared unbiased wide-field observations in the H$_2$ 1-0 S(1) line at 2.122$\mu$m alleviates the problem, enabling us to detect more outflows and trace them closer to their driving sources.} 
	{As part of a large-scale multi-waveband study of ongoing star formation in the Braid Nebula Star Formation region, we focus on a one square degree region that includes Lynds Dark Nebula\,1003 and 1004.  Our goal is to find all of the near-infrared outflows, uncover their driving sources and estimate their evolutionary phase.}
	{We use near-infrared wide-field observations obtained with WFCAM on UKIRT, in conjunction with previously-published optical and archival MM data, to search for outflows and identify their driving sources; we subsequently use colour-colour analysis to determine the evolutionary phase of each source.}
	{Within a one square degree field we have identified 37 complex MHOs, most of which are new. After combining our findings with other wide-field, multi-waveband observations of the same region we were able to discern 28 outflows and at least 18 protostars. Our analysis suggests that these protostars are younger and/or more energetic than those of the Taurus-Auriga region. The outflow data enable us to suggest connection between outflow ejection and repetitive FU\,Ori outburst events. We also find that star formation progresses from W to E across the investigated region.}
	{}  

\titlerunning{H$_2$ 2.122$\mu$m line survey of the Braid Nebula star formation region}
\authorrunning{Khanzadyan et al.}

\maketitle

\section{Introduction}

This work is part of a multi-waveband and wide-field observational campaign to study the Braid Nebula Star Formation region \citep{aspin:2009}, which is situated in the outskirts of Cygnus\,OB7. The region is part of a dark and extended ($\sim$7\degr\,$\times$\,10\degr\, in size) filamentary complex known as Kh\,141 \citep{khavtassi:1960}, or \object{TGU H541} \citep{dobashi:2005}.  It is sometimes called ``The Northern Coalsack'' and contains numerous individually identified Lynds clouds \citep{lynds:1962}. The Braid Nebula Star Formation region, defined by the \object{LDN 1003} and \object{LDN 1004} clouds, extends over a degree field and encompasses one of the highest visual extinction regions in the Kh\,141 complex. 

There have been several estimates of the distance to Cygnus\,OB7: \citet{hiltner:1956} and \citet{schmidt:1958} derived a distance of 740pc. A similar value was suggested by \citet{de-zeeuw:1999} using data from the \emph{Hipparcos} Catalogue (although their results were inconclusive due to the fact that there is a giant molecular cloud complex \citep[e.g.][]{dame:2001} at a distances of 500 to 1000pc which hampers optical population studies towards the Cygnus region). An alternative approach has been implied by \citet{pluschke:2002} using the radioactive decay of $^{26}$Al at 1.809 MeV together with the population synthesis of the OB association members and Wolf-Rayet stars in the region. Their analysis suggests a distance of 800$\pm$10pc for the Cygnus\,OB7 association, which agrees well with the earlier measurements and will therefore be assumed in our work.

\begin{figure*}
	\centering
		\includegraphics[width=18cm]{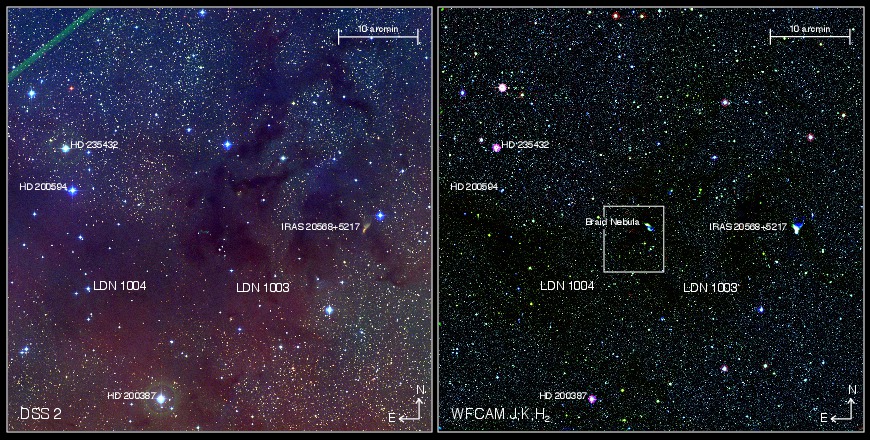}
	\caption{\textbf{Left panel:} a wide-field DSS\,2 Red image of the Braid Nebula Star Formation region. \textbf{Right panel:} a WFCAM narrow-band H$_2$ + continuum image of the same region. The box in the centre of the WFCAM image shows the extent of the area studied by \citet{movsessian:2003}. [\emph{In the online version - the left panel shows a colour composite view constructed from DSS\,2 Blue (blue), DSS\,2 Red (green) and DSS\,2 Infrared (red) images. The right panel shows a colour-composite view constructed from the WFCAM J (blue), K (green) and H$_2$ (red) bands.}]}
	\label{fig:field-overview}
\end{figure*}

The first study of the LDN\,1003 molecular cloud dates back to the discovery of a red nebulous object \object{RNO 127} by \citet{cohen:1980} which eventually turned out to be a Herbig-Haro (\object{HH 448}) object \citep{melikian:2001}. Later studies of the region uncovered numerous H$\alpha$ stars \citep{melikian:2001}, optical and near-infrared HH objects and reflection nebulae \citep{movsessian:2003}, one of which -- the Braid Nebula -- was found to be associated with an FU Orionis (FU Ori) type eruptive variable star \citep{movsessian:2006}.

Some 30$\arcmin$ west of RNO\,127, \citet{devine:1997} found several HH objects, two of which were associated with the bipolar reflection nebula created by HH\,381\,IRS (\object{IRAS 20568+5217}). This particular object did not receive attention until \citet{reipurth:1997} studied it spectroscopically and found that it has many of the characteristics of an FU Ori type star. Further spectroscopic observations supported this interpretation \citep{greene:2008} and a subsequent comparison of archival optical data from DSS-1 (1953) and DSS-2 (1990) with more recent CCD images demonstrated that HH\,381\,IRS had brightened considerably over the last few decades. The presence of two FU Ori type outburst stars in LDN\,1003, along with numerous IRAS sources \citep{movsessian:2003}, points to significant star formation activity in this area.  A follow-up unbiased, wide-field and multi-waveband study of the region was warranted; this is a project that our consortium is currently conducting.

The first stage in our exploration of the region was a near-infrared spectroscopic study of the most significant sources and nebulae \citep{aspin:2009}, where we confirmed the FU Ori nature of both the Braid Nebula Star and HH\,381\,IRS and characterised all of the observed 16 sources. Next we presented a wide-field narrowband optical survey of the region \citep{magakian:2010}, where we reported the discovery of 25 more HH objects and 12 individual outflows. The unbiased nature of the survey also enabled us to detect many reflection nebulae. In order to relate the detected outflows with the driving sources hidden behind the extended diffuse cloud filaments in LDN\,1003 and 1004 we observed the region in CSO 1.1mm continuum \citep{aspin:2011}. Our study revealed 55 cold dust clumps, some of which (20\%) coincide with IRAS sources found in the region, the rest being starless. We also found that both FU Ori objects are associated with strong millimetre emission.

In this paper we present our wide-field near-infrared H$_2$ 1-0 S(1) (2.122$\mu$m) line observations of the entire 1 square degree field centred on the Braid Nebula. Our goal is to trace the outflows and jets emanating from the protostars. Due to the significant extinction present in the region it is difficult to relate outflows detected in the optical with candidate driving sources. The influence of extinction is significantly lower in the near-infrared, which enables us to detect more outflow features. By using H$_2$ 2.122$\mu$m emission as a tracer of jets and outflows, wide-field narrow-band images may be used to pin-point the locations of the driving sources. Moreover, they allow us to take a statistical approach when considering questions about the distribution of Class 0/I protostars and Class II/III Young Stellar Objects (YSOs), the interaction of these forming stars with their surrounding environment, the overall star formation efficiency, and the evolution of the region as a whole \citep[e.g.][]{stanke:2002,davis:2007,davis:2009,kumar:2011}.

In this paper Section\,\ref{sec:obsred} describes the observations and data reduction, while Section\,\ref{sec:results} explains our results with an emphasis on the multi-waveband data used in our work. In Section\,\ref{sec:discussion} we discuss the outflows and their driving sources within the broader prospective of the entire region and use features within the outflows of the two FU\,Ori outburst stars to trace their evolution. We also search for gradients in star formation across the region. Lastly, in Section\,\ref{sec:summary} we summarise our findings and present our main conclusions.

\section{Observations and data reduction}
\label{sec:obsred}

\subsection{WFCAM data}

Wide-field near-infrared images of the Braid Nebula region were obtained in July 2006, at the United Kingdom Infrared Telescope (UKIRT\footnote{UKIRT is operated by the Joint Astronomy Centre on behalf of the Science and Technology Facilities Council of the U.K.}). The Wide-Field Camera (WFCAM) \citep{casali:2007} was used with a narrow-band H$_2$ 1-0 S(1) and broad-band J, H and K filters \citep[see][for the filter descriptions]{hewett:2006}.

WFCAM houses four Teledyne HAWAII-2 2048$\times$2048 HgCdTe detectors, each with a pixel scale of 0.4\arcsec\, that are spaced by 94\% in the focal plane. To achieve full coverage of a 54\arcmin $\times$ 54\arcmin\, area on the sky, four telescope pointings were used with a five-point jitter pattern (6.4\arcsec\, spacing) at each pointing to account for detector artefacts.  To fully sample the seeing, at each jitter position an additional 2$\times$2 micro-stepped pattern was used, shifting the array by N+1/2 pixels between each micro-step position. In the reduced images the pixel scale is therefore 0.2\arcsec . Integration times of 20 minutes and 40 minutes were reached in the broad-band and narrow-band filters, respectively.

The raw images were pipeline processed by the Cambridge Astronomical Survey Unit (CASU). The data reduction steps are described in detail by \citet{dye:2006}. All reduced WFCAM data are archived at the WFCAM Science Archive (WSA\footnote{WSA is a part of the Wide Field Astronomy Unit (WFAU) hosted by the Institute for Astronomy at the Royal Observatory, Edinburgh, UK}). Pipeline processed data were visually inspected so that residual non-uniform surface brightness variations could be removed before mosaics were constructed using the World Co-ordinate System (WCS) embedded in the headers.

The final mosaics were inspected and found to be free of any significant artefacts. Small, ring-like cross-talk features a few pixels in diameter were evident in the data adjacent to bright and saturated stars. These artefacts appear 8 and 16 pixels from each bright stars along rows and columns and were therefore easily identified and discarded during the several steps taken to detect real H$_2$ 1-0 S(1) line features. 

To identify line-emission features: \textbf{(1)} we used J, K and H$_2$ images of the same field to create a false colour RGB image in which H$_2$ features appear extremely red in colour; the cross-talk features stand-out due to their relative faintness in H$_2$. \textbf{(2)} we extracted image sections and produced continuum-subtracted data by subtracting the scaled K band images from the narrow-band H$_2$ images; compact knots remain at the locations of field stars due to imperfect subtraction, though these are easily distinguished from pure H$_2$ line-emission features, which are typically diffuse and/or extended. \textbf{(3)} each pair of H$_2$ and K band image sections were blinked together to confirm the validity of each new H$_2$ feature detection.

\subsection{Supplementary and Archival data}

In our analysis of the Braid Nebula region we compared near-infrared H$_2$ 1-0 S(1) line data with optical H$\alpha$, [SII] and CSO 1.1mm observations that were presented and thoroughly described in \citet{magakian:2010} and \citet{aspin:2011}, respectively. We have also used IRAS HiRes\footnote{HiRes uses the Maximum Correlation Method \citep[MCM,][]{aumann:1990} to produce images with better than the nominal resolution of the IRAS data.} processed data of the region.

For mid-infrared coverage AKARI/IRC data were obtained from \emph{The AKARI/IRC mid-infrared all-sky survey} catalogue \citep{ishihara:2010} where fluxes in Jy are presented for the S9W ($\lambda$ = 9$\mu$m) and L18W ($\lambda$ = 18$\mu$m) bands. We used \citet{tanabe:2008a} ZP flux densities of 56.262 and 12.001 Jy for S9W and L18W bands, respectively, to calculate the appropriate magnitudes. These are presented in the Table\,\ref{tab:photom}. Only high quality fluxes (quality flag 3) in S9W and L18W were used in our subsequent analysis of the sources in the Braid Nebula region.

The 2MASS and AKARI/IRC catalogues for the entire region were cross-correlated using a positional uncertainty of 3$\arcsec$. In this way we were able to obtain photometry in multiple bands for a near-complete population of sources in the field.

\begin{figure}
	\centering
		\includegraphics[width=8.5cm]{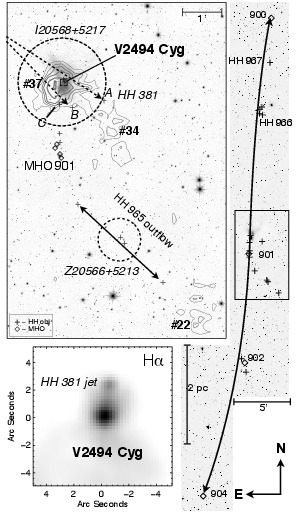}
	\caption{An H$_2$ + continuum image of the V2494\,Cyg outflow is presented in the \textbf{right, wide-field} panel. Here the diamonds and crosses indicate the positions of MHO and HH objects, respectively. The solid-line box marks the area around MHO~901 and the V2494Cyg nebula that is shown in more detail \textbf{top-left}. In this panel the grey-scale contours mark the CSO 1.1mm data with values 40, 60, 80, 100, 140 and 180 mJy; the hashed numbers label the designated CSO cores from \citet{aspin:2011}; the dashed-line circle shows the approximate extent of the IRAS HiRes 25$\mu$m cores; the dashed-line and solid-line arrows show the suspected and confirmed outflows, respectively. The \textbf{bottom-left panel} shows a magnified view of V2494\,Cyg in H$\alpha$ \citep{magakian:2010} in which the HH\,381jet can be seen.}
	\label{fig:v2494cygwide}
\end{figure}

\section{Results}
\label{sec:results}

Figure\,\ref{fig:field-overview} shows the overall extent of the Braid Nebula star formation region at optical and near-infrared wavelengths. Several known dark clouds (LDN 1003 and 1004) and bright stars, as well as the Braid Nebula and \object{IRAS 20568+5217} (also known as HH 381 IRS) are marked. Our analysis of the entire one-square-degree field reveales numerous outflow features in addition to our previous study of the region \citep{movsessian:2003}. The details of the individual \textbf{M}olecular \textbf{H}ydrogen emission-line \textbf{O}bjects (MHO\footnote{\url{http://www.jach.hawaii.edu/UKIRT/MHCat/} \citep{davis:2010}}) identified in the region are presented in the Appendix\,\ref{sec:indmho} and \ref{sec:mhodetails}.

For better identification of the protostars and the outflow driving sources in our surveyed region it is essential to probe the region at mid-infrared wavelengths.  The various photometric bands of the \emph{Spitzer}/IRAC instrument (between 3 and 25$\mu$m) are particularly sensitive to the young stellar content \citep{evans:2009}. Unfortunately, at the time of writing, there were no \emph{Spitzer} observations covering or neighbouring our field of study. This prevents us from making a detailed statistical and evolutionary study. We have therefore used mid-infrared fluxes from the AKARI/IRC catalogue. With these data we were are able to \emph{probe} the young stellar nature of the objects situated in the entire surveyed region.

To aid in our quest to identify outflow driving sources we firstly cross-correlate the available multi-waveband data for suspected candidates following the example of \citet{aspin:2009}.  These sources are presented in the Appendix\,\ref{sec:multisource}. Some of the candidates were already suggested as being young and active, based on their proximity to HH objects \citep{devine:1997,magakian:2010} and their observed spectra \citep{aspin:2009}. In this study, armed with the  locations of newly-identified near-infrared MHOs, in conjunction with the positions of recently-reported HH objects \citep{magakian:2010} and CSO 1.1mm cores \citep{aspin:2011}, we are likewise able to associate young stars with the outflows they power.  Figure\,\ref{fig:v2494cygwide} shows one example, where we confirm the 7pc bipolar outflow recently reported by \citet{magakian:2010} and are able to suggest an extension to this flow (MHO\,900 and 904), making it over 10pc long and one of the most significant outflows detected in the current region. The fact that this outflow is powered by an FU\,Ori type eruptive variable star \citep{aspin:2009,kazarovets:2011} highlights the importance of this find. Fig.\,\ref{fig:v2494cygwide} also shows the importance of using multi-waveband data; in this case, by using IRAS 25$\mu$m band data we are able to identify the source of the HH\,965 optical outflow \citep{magakian:2010}.

\begin{figure}
	\centering
		\includegraphics[width=8.5cm]{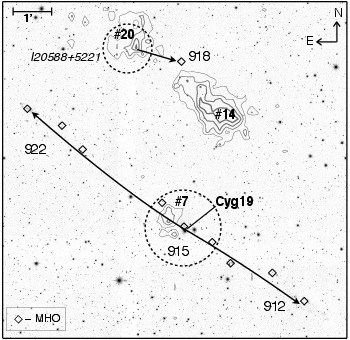}
	\caption{Similar to Fig.\,\ref{fig:v2494cygwide} but for the region around Cyg19 H$\alpha$ star where MHO 912, 915, 918 and 922 were detected.}
	\label{fig:cyg19}
\end{figure}

Figure\,\ref{fig:cyg19} reveals yet another parsec-scale outflow (MHO\,912 - 922) newly detected in near-infrared H$_2$ 1-0 S(1) emission. There are three CSO 1.1mm cores in this sub-field (\#7, \#14 and \#20). Two seem to be associated with outflows: MHO 912-922 and MHO\,918. Surprisingly, the brightest and most compact core (\#14) is not obviously associated with an MHO or HH outflow. Core \#7 is possibly associated with the H$\alpha$ emission-line star Cyg\,19 \citep{melikian:1996}, which we identify as the driving source of this bipolar outflow.

\begin{figure}
	\centering
		\includegraphics[width=8.5cm]{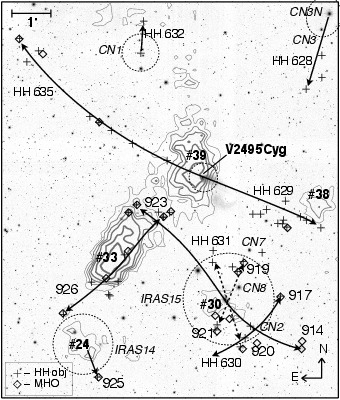}
	\caption{Similar to Fig.\,\ref{fig:v2494cygwide} but for the region around the Braid Nebula where numerous HHs and MHOs were detected. Solid-line and dashed-line arrows show confirmed and suspected outflows, respectively. Cometary Nebulae (CN) reported by \citet{magakian:2010} are marked for reference.}
	\label{fig:braid}
\end{figure}

\begin{table*}
	\caption[]{\label{tab:outflows}List of outflows and their properties}
	\centering
	\begin{tabular}{lccccc}
		\hline\hline
	Driving
		& Outflow
		& \multicolumn{2}{c}{Projected\tablefootmark{a}}
		& P.A.\tablefootmark{b}
		& $L_{2.12}$\tablefootmark{c}
		\\
	Source
		& Components
		& \multicolumn{2}{c}{Length (pc)}
		& (degs)
		& $\times 10^{-3}L_{\sun}$
		\\
		\hline
Z20566+5213
	& HH 695NE knot+bow, HH 965SW knot+bow	
	& 0.35
	& 0.33	
	& 47$\pm$5	
	& \ldots
\\
V2494 Cyg	
	& MHO 900, HH 967, HH 966, HH 381jet, MHO 901, MHO 902, MHO 904
	& 4.67 
	& 5.60	
	& 174$\pm$5	
	& 28.77
\\
I20575+5221	
	& MHO 906A, B and C, HH 968
	& 1.06 
	& 0.35 
	& 38$\pm$5
	& 9.17
\\
core \#44
	& MHO 907, MHO 909, MHO 910
	& 0.70 
	& 0.70
	& 10$\pm$5
	& 59.11
\\
I20575+5210 
	& MHO 908, HH 974, MHO 903
	& 0.23 
	& 0.69
	& 95$\pm$5
	& 61.63
\\
I20575+5210
	& HH 971, HH 972, HH 973
	& 0.56 
	& 0.29
	& 0$\pm$5
	& \ldots
\\
CN6 (core \#51)
	& MHO 911 (HH 627A)
	& 0.21 
	& \ldots
	& 140$\pm$5
	& 28.32
\\
R20582+5221
	& HH 627C, HH 627D
	& 0.19 
	& \ldots
	& 228$\pm$5
	& \ldots
\\
I20583+5228
	& MHO 913A, B and C
	& 0.10 
	& 0.10
	& 150$\pm$5
	& 7.14
\\
CN3N
	& HH 628
	& 0.48 
	& \ldots
	& 165$\pm$5
	& \ldots
\\
Cyg19
	& MHO 912, MHO 915, MHO 922
	& 1.02 
	& 1.27
	& 54$\pm$5
	& 109.19
\\
CN2
	& MHO 917, HH 630A
	& 0.26 
	& 0.14
	& 129$\pm$5
	& 3.28
\\
IRAS 15N
	& MHO 914, HH 976knot, MHO 923C and D
	& 0.55 
	& 0.83
	& 30$\pm$5
	& 56.28
\\
I20588+5221
	& MHO 918
	& 0.29 
	& \ldots
	& 252$\pm$5
	& 1.70
\\
V2495 Cyg
	& HH 629, MHO 916, MHO 924, MHO 927
	& 0.76 
	& 1.37
	& 55$\pm$5
	& 15.46
\\
CN1 
	& HH 632jet+knot A
	& 0.20 
	& \ldots
	& 0$\pm$5
	& \ldots
\\
core \#33
	& MHO 923A and B, MHO 926
	& 0.55 
	& 0.55
	& 135$\pm$5
	& 44.74
\\
IRAS 14
	& MHO 925
	& 0.19 
	& \ldots
	& 185$\pm$5
	& 3.15
\\
I21005+5217
	& MHO 928, MHO 929, MHO 933
	& 0.28 
	& 0.56
	& 130$\pm$25
	& 11.73
\\
I21005+5217
	& MHO 930, MHO 935
	& 0.20 
	& 0.46
	& 35$\pm$5
	& 10.75
\\
\\
\multicolumn{6}{c}{Suspected Outflows}\\
\hline
\ldots
	& MHO 905
	& \ldots
	& \ldots
	& 45$\pm$15
	& 1.06
\\
core \#44(?)
	& HH 969
	& \emph{0.35}
	& \ldots
	& 280$\pm$15
	& \ldots
\\
IRAS 15N
	& MHO 919, MHO 921, HH 977
	& 0.25 
	& 0.20
	& 160$\pm$5
	& 15.89
\\
IRAS 15N
	& MHO 920, MHO 921C, HH 631A and B
	& 0.27 
	& 0.22
	& 9$\pm$5
	& 11.37
\\
core \#26
	& MHO 936
	& 0.19 
	& \ldots
	& 219$\pm$15
	& 13.17
\\
core \#15
	& MHO 931
	& 1.15 
	& \ldots
	& 127$\pm$5
	& 6.09
\\
core \#8
	& MHO 932
	& 0.70 
	& \ldots
	& 70$\pm$15
	& 28.12
\\	
core \#40
	& MHO 934
	& 1.74 
	& \ldots
	& 310$\pm$15
	& 3.05
\\
	\hline
	\end{tabular}
	\tablefoot{A table presenting a list of outflows detected in our studied region with their characteristics (cf. Sect.\,\ref{sec:results}).
	\tablefoottext{a}{Sizes of two separate lobes (if observed) calculated assuming a distance of 800~pc to the region.}
	\tablefoottext{b}{Position Angle of the outflow (with error) measured East of North.}
	\tablefoottext{c}{The luminosity is calculated by summing the de-reddened H$_2$ 1-0 S(1) line flux values for the individual components presented in Table\,\ref{tab:mhos} and converting these fluxes into a luminosity assuming a distance of 800 pc.}
	}
\end{table*}

Figure\,\ref{fig:braid} shows the immediate surroundings of the Braid Nebula. There is a considerable number of HH objects and MHOs in this region. Identifying the driving sources of each outflow is difficult; spectroscopic studies would help narrow down the number of YSO candidates in the region. Here we have yet another case of FU\,Ori type outburst star, V2495\,Cyg \citep{movsessian:2006,kazarovets:2011}, which illuminates the Braid Nebula and drives the parsec-scale bipolar outflow (V2495\,Cyg outflow). This region hosts a number of prominent IRAS objects (dashed circles) and CSO 1.1mm cores (solid-line contours) which most likely host several protostars.

A complete and detailed description of the other regions where outflows are detected is presented in Appendix\,\ref{sec:outfldet} and \ref{sec:mhodetails}; in these appendices each MHO and its association with neighbouring MHOs and YSOs is described in detail. Here in Table\,\ref{tab:outflows} we present only an overview of each region, where the first column lists the driving sources followed by the names of individual components in the second column. The third and fourth columns give separate lobe lengths for each outflow in parsecs. The fifth column lists the outflow orientation measured East of North in degrees, while the sixth column lists the H$_2$ 1-0 S(1) line luminosity in units of L$_{\sun}$ for the entire outflow. The luminosity was calculated by combining the extinction corrected H$_2$ 1-0 S(1) luminosity values for each individual MHO in the outflow. The extinction values used in the calculation (see Tab.\,\ref{tab:mhos} for more details) were obtained from the near-infrared extinction map discussed later in (Sec.\,\ref{sec:discussion}).

\begin{figure*}
	\centering
		\includegraphics[width=14cm]{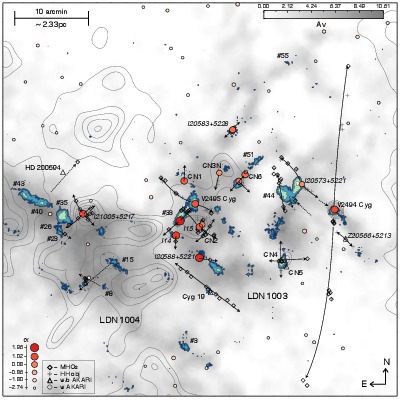}
	\caption{An A$_V$ map of the surveyed region constructed from WFCAM J, H and K band data using the technique described in \citet{rowles:2009}. The overlaid grey-scaled contours are from IRAS HiRes 100$\mu$m data; these increase from 10\% of the local field maximum in steps of 10\%. Filled contours represent the CSO 1.1mm data with values of 40, 60, 100 and 180 mJy. The hashed numbers are the designated CSO core numbers from \citet{aspin:2011}. Sources described and discussed in Sect.\,\ref{sec:protcont} are presented using the same symbols as in Fig.\,\ref{fig:ccdjhhk}. The sources with the most positive values of $\alpha$ are marked with a larger and darker-coloured symbol; the objects with the most negative values of $\alpha$ are marked with a smaller and lighter-coloured symbol. The outflows listed in Tab.\,\ref{tab:outflows} are indicated with solid-line arrows; suspected flows are marked with dashed-line arrows.}
	\label{fig:outflows}
\end{figure*}

The observations described above reveal the extent of outflow activity in this region.  
At least 28 outflows are identified (Tab.\,\ref{tab:outflows}) of which 15 are bipolar and 5 monopolar. There are eight additional cases (two bipolar and six monopolar) where there is some ambiguity and further studies are essential. The two unconfirmed bipolar outflows possibly emanating from IRAS\,15N (see Sect.\,\ref{sec:braid}) are difficult to confirm without further high-resolution imaging and spectroscopy due to the very complicated nature of the surrounding environment (Fig.\,\ref{fig:braid}). Finally, we suggest that the locations of the MHO\,931 and 932 driving sources are coincident with cores \#15 and \#8, respectively (Figure\,\ref{fig:outflows}), while the bow-shock MHO\,934 is probably driven by core \#40 or even core \#43 (but obviously not by HD\,200594).

\section{Discussion}
\label{sec:discussion}

\subsection{Outflow parameters}
\label{sec:outflparam}

Figure\,\ref{fig:outflows} reveals the map of visual extinction (A$_V$) for the entire surveyed field, where the positions of MHOs and HH objects are marked and the directions of associated outflows indicated. YSO positions are overlaid together with CSO 1.1mm and IRAS HiRes contours. The outflow features are generally confined to the main high-extinction regions, namely LDN\,1003 and 1004, where most of the molecular material is located. The combined representation of A$_V$ and CSO 1.1mm illustrates all of the dust contained in the region, since for the higher extinction regions the A$_V$ map is less sensitive to high column density structures, in contrast to the CSO 1.1mm map which is more sensitive in these regions \citep{kirk:2006,davis:2008}.

\begin{figure}
	\centering
		\includegraphics[width=8.8cm]{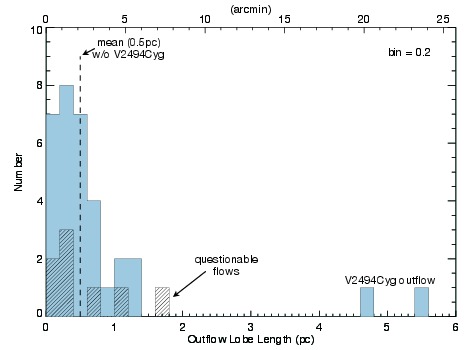}
	\caption{Histogram of the outflow lobe lengths, measured in parsecs and arc-minutes, as listed in Tab.\,\ref{tab:outflows}. The filled histogram bars exclude the questionable outflows (identified in Tab.\,\ref{tab:outflows}), which are represented by the hatched bars.}
	\label{fig:outflowhist}
\end{figure}

Figure\,\ref{fig:outflowhist} shows the distribution of outflow lobe lengths uncorrected for flow inclination angle. We observe a range of flow lengths, from 0.1~pc (I20583+5228) to 1.37~pc (V2495\,Cyg). The distribution peaks at around 0.3~pc. The mean value of the distribution without the V2494\,Cyg outflow components, which differs substantially from the majority of flows detected in the region, is about 0.5~pc. If we assume a flow inclination angle of 45 degrees to the line of sight, then the mean outflow single-lobe length is 0.71~pc. It would take at least 13,800 years for an outflow with a constant speed of 50 km s$^{-1}$ to propagate over this distance, assuming a uniform density medium \citep{arce:2007}. This mean dynamical age ($\sim$10$^4$ yr) for the outflows in the Braid Nebula region is comparable to the duration of the low-mass protostellar phase, as derived from statistical studies of the protostellar populations in a number star formation regions. These studies indicate a protostellar lifetime of 10$^4$$-$10$^5$ years \citep{froebrich:2006,hatchell:2007,enoch:2008,davis:2009}. Our estimate of the mean flow lifetime also indicates that, in the Braid Nebula Region, we are observing a recent phase of star formation.

\citet{evans:2009} predict a longer time-span for the protostellar phase, of the order of 10$^5$$-$10$^6$ years. This may indicate that most of our outflows are driven by very early phase (Class\,0 or Class I) protostars.  However, as we discuss later, this may not be the case.  Instead, we may be underestimating the mean flow dynamical age; we may not be detecting the terminating shocks of the outflows when they leave the main high-extinction ridge-like structure associated with the Braid Nebula region (Fig.\,\ref{fig:outflows}) and therefore may be underestimating the mean flow length. In reality, the apparent filamentary structure of the molecular cloud in Fig.\,\ref{fig:outflows} suggests a more clumpy environment, which would result in the detection of outflow features (HH objects and MHOs) only when there is sufficiently dense ambient molecular material to interact with. Such a scenario has been reported in several similar wide-field studies: of Corona Australis \citep{kumar:2011}, Orion \citep{davis:2009} and the DR21 region \citep{davis:2007}. Furthermore, the distance to the region would impose a lower limit for the detected outflow lengths, preventing a proper statistical analysis. Therefore any attempt to characterise the outflow lengths should be interpreted with caution.

\begin{figure}
	\centering
		\includegraphics[width=8.8cm]{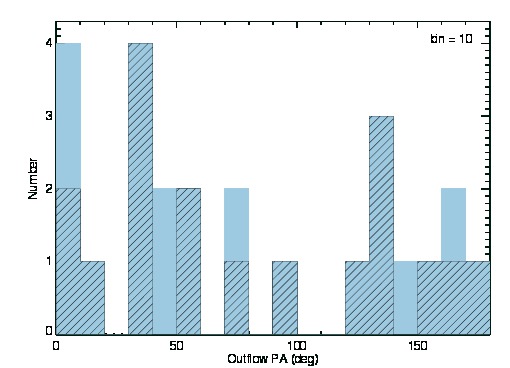}
	\caption{Distribution of outflow position angles (P.A.) measured East of North (see Tab.\,\ref{tab:outflows}). The filled histogram bars shows all of the detected outflows; the hatched bars show only the bipolar outflows.}
	\label{fig:outflow-angle}
\end{figure}

For outflow orientation angles (Tab.\,\ref{tab:outflows}) we performed a Kolmogorov-Smirnov (KS) test which gave us 95$\%$ probability that our distribution is drawn from a homogeneously distributed sample. Any further attempts to get a correlation between outflow position angles and the orientations of the long axes of associated IRAS sources or CSO 1.1mm cores was thought to be premature at this stage, due to the low statistical sample in our possession. However, our current analysis does seem to suggest that there is no preferable outflow orientation in the region (Fig.\,\ref{fig:outflow-angle}), as is found to be the case in many other regions \citep[e.g. Orion: ][]{davis:2009}.

Table\,\ref{tab:outflows} also lists the luminosity of the outflows in H$_2$ 1-0 S(1) line emission, corrected for the extinction. The missing values correspond to the few cases (six out of 28) when outflows were only detected at optical wavelengths. These flows could produce H$_2$ 1-0 S(1) line emission that is fainter than our detection limit of $\sim$10$^{-7}$ W m$^{-2}$ calculated in a standard 5\arcsec aperture, which would corresponds to $\sim$ 0.2$\times$ 10$^{-3}$ L$\sun$ per outflow component (without an extinction correction). In a few cases (eight out of the 22 remaining) outflows are only partially detected in H$_2$ 1-0 S(1) line: for example, sometimes one of the bipolar outflow lobes is traced by optical HH objects, without any corresponding near-infrared shock features (MHOs). In these cases we still present the integrated luminosities of one lobe (if it is detected in the NIR) in Tab.\,\ref{tab:outflows}, but regard these values as a lower limit in our subsequent determination of the outflow power necessary to drive these shocks using bow shock models \citep[cf.][]{smith:1991,smith:2003} or hydrodynamic simulations \citep[cf.][]{smith:2005,smith:2007}.

As a comparison, the outflows detected in Orion \citep{stanke:2002}, which is half the distance to the Braid Nebula region, range from 0.04 to 20 $\times$ 10$^{-3}$ L$\sun$, while the faintest outflows detected in DR\,21 \citep{davis:2007}, which is about four times further than our region, have luminosities of about 30 $\times$ 10$^{-3}$ L$\sun$ in H$_2$ 1-0 S(1) line. Sensitivity limitations affect the completeness of statistical samples of outflows, since we tend to detect only the luminous ones as we observe more distant regions.

The force of the outflows from protostars (Class\,0/I sources) is correlated with the bolometric luminosity over several orders of magnitude \citep{beuther:2002}. For protostars, the bolometric luminosity (L$_{bol}$) is dominated by the energy released through accretion, expressed as the accretion luminosity ($L_{acc}$).  In order to reliably determine or give an estimate of this value, more thorough modelling is required, based on a well sampled Spectral Energy Distribution (SED) for each source. Unfortunately, our current SED coverage for the driving source candidates in the Braid Nebula Region is quite limited (see Tab.\,\ref{tab:photom}). This prevents us from obtaining reliable estimates for source bolometric luminosities. However, as an approximation, we could connect the outflow power to the $L_{bol}$ of the driving source, using the known correlation with outflow force (and a weak velocity dependence).

However, the accretion rate also depends on the evolutionary phase. For low-mass stars, the most luminous outflows are generally driven from the youngest (Class\,0/I) protostars in the same mass range \citep{stanke:2003} due to the high early accretion rate. The outflow power and the observed luminosity would then decrease with the evolutionary age due to the diminishing accretion rate. However, this relation is also strongly dependent on the mass of the protostar itself, with the more massive sources possibly increasing their accretion rates as they evolve through these early stages \citep[e.g. the accelerated accretion model of][]{mckee:2003}. In this case, luminous outflows  could be detected even from the relatively evolved massive protostellar sources.

\begin{figure}
	\centering
		\includegraphics[width=8.8cm]{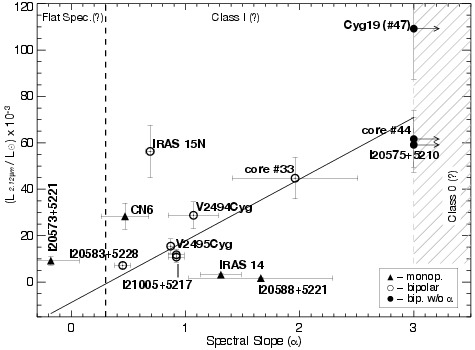}
	\caption{A plot showing the dependence of the near- to mid-infrared spectral slope ($\alpha$) to the luminosity of the associated outflow derived from the H$_2$ 1-0 S(1) line flux. The vertical hatched area around $\alpha$ $=$ 3 shows the upper tentative border for the Class\,I and 0 protostars according to \citet{lada:1987}; the vertical dashed line at $\alpha$~$=$~0.3 indicates the border between Class\,I and Flat Spectra type sources according to \citet{greene:1994}. The error for the flux values was assumed to be of the order of 20$\%$. The solid line represents the derived best-fit for the plotted data points of the form: $y=(26.5\pm2.9)x - (8.8\pm4.6)$ with $\chi^2$ = 77.6.}
	\label{fig:alpha-h2}
\end{figure}

Some of the known outflows from massive protostars have substantially higher luminosities than their low-mass counterparts. For example, IRAS\,05358+3543 \citep{beuther:2002a} drives the N1-N6 outflow (MHO 1401-1406) which in H$_2$ 1-0 S(1) line emission has a luminosity of 136 $\times$ 10$^{-3}$ L$\sun$ \citep{khanzadyan:2004}, while the IRAS\,20126+4104 jet MHO 860 \citep{caratti-o-garatti:2008} produces an H$_2$ 1-0 S(1) line luminosity of 128 $\times$ 10$^{-3}$ L$\sun$. By comparison, the well known low-mass Class\,0 protostellar outflows, HH\,212 and HH\,211, have luminosities in H$_2$ 1-0 S(1) emission of 6.9 and 3.4 $\times$ 10$^{-3}$ L$\sun$, respectively \citep{stanke:2002}. The outflows in the Braid Nebula region on average have higher luminosities than would be expected from low-mass protostars (Tab.\,\ref{tab:outflows}).  This would suggest that our outflows are being driven by at least intermediate mass or possible even high mass protostars.

Lastly, in Fig.\,\ref{fig:alpha-h2} we look for a relationship between the H$_2$ 1-0 S(1) line luminosity of the flows (Tab.\,\ref{tab:outflows}) with the near- to mid-infrared spectral slope ($\alpha$) of the driving source (Tab.\,\ref{tab:ccanalysis}).  Discussion of source spectral slopes will be presented in Sect.\,\ref{sec:protcont}; here we simply assume that $\alpha$ is an indication of the age of the source, where higher values correspond to younger driving sources (simply because there is more dust around the source). The original scheme described in \citep{greene:1994} regards values between -0.3 and 0.3 as flat spectrum sources and values greater than 0.3 as Class\,I sources. Since Class\,0 sources (which are thought to be younger than Class I objects) do not have any emission in the mid-infrared, we assume a tentative value of $\alpha \sim 3$ for the border between Class 0 and Class I objects \citep{lada:1987}, the latter having  $\alpha$ $<$ 3.

The best-fit line shown on Fig.\,\ref{fig:alpha-h2} indicates a dependency between outflow power and the evolutionary state of the driving source. The younger the system is the higher the overall outflow luminosity in H$_2$ 1-0 S(1) line emission, which means the more energetic and more powerful the outflow is. The power of the outflows, on the other hand, decreases as the protostar ages which would seem to be a logical conclusion since a similar pattern is observed in low-mass outflows \citep{stanke:2003}. However, the fit to the data in Fig.\,\ref{fig:alpha-h2} is relatively poor. This could be due to the different masses and multiplicity of the driving sources. There are several other factors which could have an influence on the data: erroneous extinction calculations towards the outflow components, the somewhat uncertain distance estimate to the region, and questionable spectral slope ($\alpha$) determinations using the AKARI bands.

A more comprehensive approach to the outflow luminosity analysis is essential to test the relationship between protostellar evolutionary phase and outflow flux. In particular, a correlation between outflow luminosity and the bolometric luminosity, $L_{bol}$, of the driving source may exist. Such a correlation has been demonstrated for low-mass protostars and their outflows \citep{caratti-o-garatti:2008}; does this relationship extend towards intermediate and high-mass protostars? Unfortunately the current limited dataset for the driving sources in the Braid Nebula region (Tab.\,\ref{tab:sources}) limits the accuracy of our $L_{bol}$ determinations, although additional observing campaigns are underway to fill the existing gaps in the source SEDs.

\subsection{Protostellar content}
\label{sec:protcont}

\begin{figure}
	\centering
		\includegraphics[width=8.8cm]{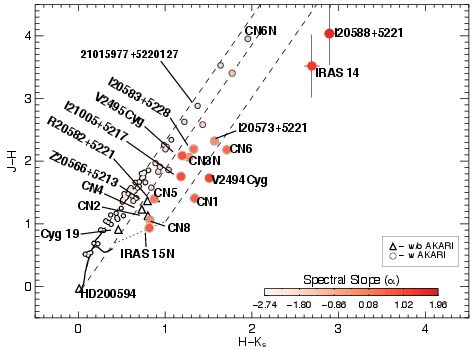}
	\caption{A near-infrared J - H\,vs.\,H - K$_S$ colour-colour diagram for sources in the studied field with high quality ($q=3$) AKARI/IRC fluxes. Values are plotted in the 2MASS photometric system; the main sequence, dwarf and giant loci, as well as the CTTS loci \citep{meyer:1997}, are drawn along with reddening vectors. The colour bar and the size of the circles indicate the value of spectral slope ($\alpha$) calculated over the K to 9$\mu$m or K to 18$\mu$m range of the SED (see Sect.\,\ref{sec:protcont} for an explanation).}
	\label{fig:ccdjhhk}
\end{figure}

\begin{figure}
	\centering
		\includegraphics[width=8.8cm]{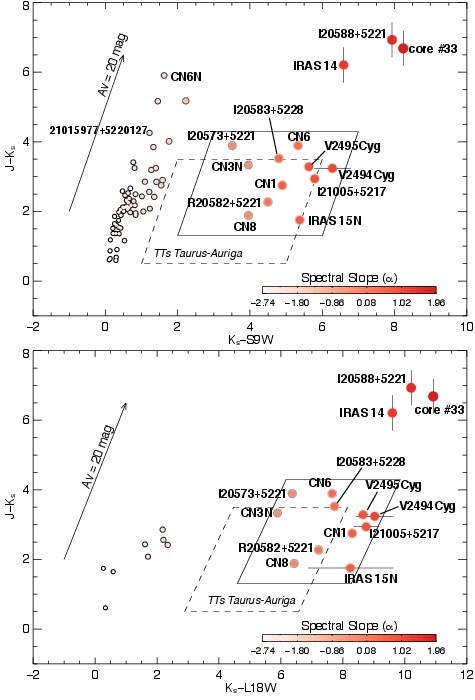}
	\caption{Near- to mid-infrared colour-colour diagrams for the same sources as plotted in Fig.\,\ref{fig:ccdjhhk}. \textbf{Top panel:} a J~-~K$_S$\,vs.\,K$_S$~-~S9W colour-colour diagram  \citep{takita:2010} for all sources found in the one-square-degree field with valid near-IR fluxes in the 2MASS catalogue and 9$\mu$m fluxes in the AKARI/IRC catalogue (Fig.\,\ref{fig:field-overview}). The arrow shows the interstellar extinction vector estimated from the \citet{weingartner:2001} Milky Way model for R$_V$ = 3.1. Dashed-line and solid-line parallelograms show the distribution of CTTSs in the Taurus-Auriga region \citep{takita:2010} and the distribution of driving sources in our region, respectively. \textbf{Bottom panel:} a J - K$_S$\,vs.\,K$_S$ - L18W colour-colour diagram similar to the top panel, but for sources with valid JHK$_S$ fluxes in the 2MASS catalogue and 18$\mu$m fluxes in the AKARI/IRC catalogue. Colour-bars and symbol sizes indicate the magnitude of the spectral slope ($\alpha$).}
	\label{fig:ccdmir}
\end{figure}

A first look at the multi-waveband source association (Tab.\,\ref{tab:sources}) reaffirms some of the suggestions for the driving sources (16 out of 24) due to their detections in the mid-infrared range which is indicative of a circumstellar disk. In relatively few cases (14 objects) the driving source candidates are detected in the far-infrared to millimetre range which tends to be when the source is embedded and young. Core\,\#44 and IRAS\,20575+5210 sources are only detectable in the far-infrared to millimetre range which is indicative of a very young, probably Class\,0 stage protostar pending further confirmation.

In order to determine the protostellar content of the studied region we cross-correlated the near-infrared 2MASS K$_s$ and mid-infrared AKARI/IRC (9 and 18$\mu$m) catalogues using only the high quality photometric measurements. This resulted in the detection of 76 sources with both K$_s$ and S9W band measurements. Of these 76 sources, only 22 have L18W band data. We performed the same procedure using our WFCAM K band photometric catalogue\footnote{The detailed description of our WFCAM near-infrared broad-band dataset will be presented elsewhere, but for the current work we used the catalogue in a few cases to obtain better estimates of the near-infrared photometric measurements or limiting magnitudes for J and H bands.} and obtained identical results because of the overall low sensitivity of the AKARI/IRC measurements. Finally, we have also included some additional sources which have no AKARI/IRC match but are suspected of being a driving source due to their locations relative to HH objects and MHOs.

In Figure\,\ref{fig:ccdjhhk} we present a near-infrared J\,- H\,vs.\,H\,-\,K$_s$ colour-colour diagram where only the outflow driving sources and the suspected candidates are marked for clarity. The majority of the sources appear to be relatively reddened main-sequence stars. Most of the outflow driving sources exhibit typical CTTS colour characteristics with some amount of extinction \citep{meyer:1997}. Some sources, like IRAS\,14, IRAS\,20588+5221 and core\,\#33 (not shown due to H\,-\,K$_s$=5.95) stand-out as being intrinsically red. 

In Figure\,\ref{fig:ccdmir}(top) we show a near- to mid-infrared J\,-\,K$_s$\,vs.\,K$_s$\,-\,S9W colour-colour diagram for the same sources plotted in Figure\,\ref{fig:ccdjhhk}. Here one can clearly see the main-sequence stars which occupy a distinct region to the left of the plot, as noted by \citet{takita:2010}. The population of outflow driving sources in the Braid Nebula region are enclosed by a parallelogram; this parallelogram is clearly shifted with respect to the CTTS stars observed in the Taurus-Auriga complex, which are indicated by the dashed-line parallelogram. Although the apparent shift in the average colours of the Braid nebula sources could be due to increased extinction along the line of sight to the region (the Braid is more distant than Taurus-Auriga, by a factor of $\sim$2), many of the sources are redder in K$_s$-S9W and K$_s$-L18W. The position of the dashed-line parallelogram on the diagram suggests the presence of excess radiation at or around 9$\mu$m, which is generally consistent with the presence of accreting disk. Three of the driving sources (IRAS\,14, IRAS\,20588+5221 and core\,\#33) appear extremely red on the J\,-\,K\,vs.\,K\,-\,S9W diagram.  These three sources may be particularly young, or they may simply be viewed ``edge-on'' (i.e. with their outflow axes close to the plane of the sky). 

Fig.\,\ref{fig:ccdmir} (bottom) shows a J\,-\,K\,vs.\,K\,-\,L18W colour-colour diagram. Only sources with a clear AKARI/IRC 18$\mu$m detection are plotted.  This diagram further highlights the reddening and extreme youth of IRAS\,14, IRAS\,20588+5221 and core\,\#33. The excess radiation at mid-IR wavelengths from their protostellar envelopes is particularly strong. The shifted between the Taurus-Auriga CTTSs and the Braid Nebula region driving sources is also more extreme in this figure.

Our colour-colour analysis of the sources in the studied region reveals the apparent protostellar nature of the driving sources. It also appears to show the deficit of CTTSs in the region. However, in our view this is a result of observational bias. The limited sensitivity of the AKARI/IRC data (0.05Jy in 9$\mu$m and 0.09Jy in 18$\mu$m) and the larger interstellar extinction towards the Braid Nebula would prevent the detection of many of the CTTSs, which in general would have a mid-infrared flux that is below our detection limit. Alternatively, this could indicate that Braid Nebula Star Formation region hosts more energetic (two FU\,Ori type stars) and relatively massive protostars than that of the Taurus-Auriga system. 

\begin{figure}
	\centering
		\includegraphics[width=8.8cm]{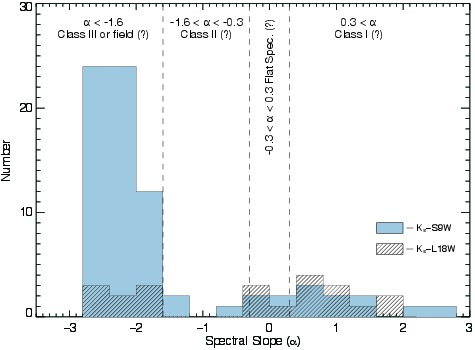}
	\caption{A histogram showing the distribution of spectral slope ($\alpha$) for the sources with valid fluxes in the 2MASS and AKARI/IRC catalogues for the entire one-square-degree field. Slopes derived using K to 9$\mu$m photometry and K to 18$\mu$m data are plotted separately using filled and hatched histogram bars, respectively. Protostellar classifications are also indicated following \citet{greene:1994} and \citet{evans:2009}.}
	\label{fig:ccdhist}
\end{figure}

\begin{table*}
	\centering
	\caption{Information pertaining to the evolutionary stage of the outflow source candidates\label{tab:ccanalysis}}
	\begin{tabular}{lccccccc}
	\hline\hline
Source & RA & Dec & $\alpha_{18}$\tablefootmark{a} & $\alpha_{9}$\tablefootmark{a} & & Outfl. & \\
Name   & (2000.0) & (2000.0) & (K-L18W) & (K-S9W) &Class\tablefootmark{b} & Det. & Comment \\
	\hline
Z20566+5213 
	& 20:58:11.41	& +52:25:27.9	& \ldots	& \ldots	& \ldots	&	Y 	& no AKARI\\
V2494Cyg
	& 20:58:21.10	& +52:29:27.9	& 1.07($\pm$0.22)	& 1.32   	& I			& 	Y	&		\\
I20573+5221
	& 20:58:50.09	& +52:32:54.3	& -0.18($\pm$0.25)	& -0.47		& F			& 	Y	&		\\
CN5
	& 20:59:04.19	& +52:21:44.8	& \ldots	& \ldots	& \ldots	& 	N	&no AKARI\\
core \#44
	& 20:59:04.72	& +52:31:38.2	& \ldots	& \ldots	& 0?		& 	Y	&no NIR+MIR\\
I20575+5210
	& 20:59:07.63 	& +52:22:21.7	& \ldots	& \ldots	& 0?		& 	Y	&no NIR+MIR\\
CN4
	& 20:59:08.95	& +52:22:39.2	& \ldots	& \ldots	& \ldots	& 	N	&no AKARI\\
CN6N
	& 20:59:40.01	& +52:34:36.2	& \ldots	& -1.68		& III?		& 	N	&field\\
CN6(core \#51)
	& 20:59:40.72	& +52:34:13.6	& 0.47($\pm$0.21)	& 0.71		& I			& 	Y	&\\
R20582+5221
	& 20 59 46.20	& +52 33 21.6	& 0.23($\pm$0.05)& 0.17		& F			& 	Y	&\\
I20583+5228
	& 20:59:51.85	& +52:40:20.6	& 0.45($\pm$0.07)& 0.36		& I			& 	Y	&\\
CN3N
	& 21:00:03.77	& +52:34:29.0	& -0.31($\pm$0.12)& -0.18		& II		& 	Y	&\\
Cyg 19
	& 21:00:11.59	& +52:18:17.2	& \ldots	& \ldots	& 0?		& 	Y	&no AKARI\\
CN2
	& 21:00:16.61	& +52:26:22.9	& \ldots 	& \ldots 	& \ldots	& 	Y	&no AKARI\\
CN8
	& 21:00:19.04	& +52:27:28.2	& -0.11($\pm$0.06)& -0.18		& F			& 	?	&\\
IRAS 15N
	& 21:00:21.41	& +52:27:09.5	& 0.69($\pm$0.04)& 0.74		& I			& 	Y	&\\
I20588+5221
	& 21:00:21.43	& +52:22:57.1	& 1.66($\pm$0.63)& 2.38		& I			& 	Y	&\\
V2495Cyg
	& 21:00:25.38	& +52:30:15.5	& 0.87($\pm$0.03)& 0.90		& I			& 	Y	&\\
CN1
	& 21:00:35.30	& +52:33:26.0	& 0.67($\pm$0.21)& 0.42		& I			& 	Y	&\\
core \#33
	& 21:00:38.80	& +52:27:58.0	& 1.96($\pm$0.55)& 2.58		& I			& 	Y	&\\
IRAS 14
	& 21:00:42.38	& +52:26:00.7	& 1.31($\pm$0.18)& 1.52		& I			& 	Y	&\\
21015977+5220127
	& 21:01:59.77	& +52:20:12.7	& \ldots	& -1.95		& III?		& 	N	&field\\
I21005+5217
	& 21:02:05.47	& +52:28:54.4	& 0.92($\pm$0.07)& 1.00		& I			& 	Y	&\\
HD 200594
	& 21:02:23.91	& +52:34:32.9	& \ldots	& \ldots	& \ldots	& 	N	&no AKARI\\
	\hline
	\end{tabular}

	\tablefoot{This table lists the evolutionary phase for the sources highlighted in Figs.\,\ref{fig:ccdjhhk} and \ref{fig:ccdmir}. The classification follows closely the \citet{greene:1994} scheme where Class\,\textbf{III} sources have $\alpha$ $<$ -1.6, Class\,\textbf{II} have -1.6 $\lesssim$ $\alpha$ $<$ -0.3, \textbf{F} or Flat Spectrum sources -0.3 $\lesssim$ $\alpha$ $<$ 0.3 and Class\,\textbf{I} sources have 0.3 $\lesssim$ $\alpha$. \tablefoottext{a}{Spectral Indices are calculated between K and 9$\mu$m and, where possible, between K and 18$\mu$m. Errors are presented for the former only since between K and 9$\mu$m a simple linear fit is applied.}
\tablefoottext{b}{The spectral class is derived from $\alpha_{18}$; $\alpha_{9}$ is used if 18$\mu$m AKARI data are not available.}
}
\end{table*}

The Spectral Index ($\alpha$), the slope of the SED calculated between 2 to 20$\mu$m, is another useful method for determining the protostellar evolutionary phase \citep{lada:1987}. This method has been revised and adjusted several times \citep{greene:1994} to explain some of the new details emerging from recent multi-waveband large scale studies \citep[e.g.][]{evans:2009}. Here we use the same approach \citep{greene:1994} and calculate the spectral slope of the SED for the sources in the studied region. Figure\,\ref{fig:ccdhist} shows a histogram of near- to mid-infrared Spectral Slope values for all of the sources found in the Braid Nebula region with valid 2MASS and AKARI/IRC fluxes. The filled histogram bars (Fig.\,\ref{fig:ccdhist}) show the distribution of sources with $\alpha$ calculated between K and 9$\mu$m; the hatched histogram bars shows the distribution for sources with values calculated between K and 18$\mu$m. The vertical dashed lines mark the ranges in $\alpha$ that correspond to the various protostellar phases \citep{greene:1994}. This plot also aids in distinguishing the protostellar content of the Braid Nebula Star Formation Region. We see two separate distributions, one tracing the younger stellar content (-1 $<$ $\alpha$ $<$ 3) and the other the more evolved or most likely background objects (-1 $<$ $\alpha$ $<$ -3).

In Table\,\ref{tab:ccanalysis} we list the driving source candidates with their coordinates along with values for the fitted Spectral Index ($\alpha$) measured between K and 18$\mu$m ($\alpha_{18}$) and K and 9$\mu$m ($\alpha_{9}$). We also list the associated source classification using $\alpha_{18}$ following \citet{greene:1994}. It is worth noting here that this classification scheme has never been tested with AKARI/IRC fluxes and consequently the reliability of the procedure is unknown. The value of $\alpha$ can be different when one of the AKARI/IRC fluxes is missing. In such cases the source could be miss-classified, so the current results should be used with some degree of caution and, if possible, with the positional information of the sources with respect to dense cores in the region. Never-the-less, the statistics suggest that most molecular outflows are associated with Class\,I protostars, in agreement with many similar studies \citep[e.g. Orion][]{davis:2009}. Further multi-waveband analysis and SED modelling will be carried-out in our continued study of these sources to determine the exact parameters and the evolutionary stage of the protostars in the studied region.

\subsection{Outflows from FU Ori stars}

The Braid Nebula Region contains two FU\,Ori type outburst stars \citep{aspin:2009}, V2494\,Cyg and V2495\,Cyg, both of which drive parsec-scale outflows traced by chains of HH objects and MHOs (Table\,\ref{tab:outflows}). The association of outflows with FU\,Ori type outburst stars is not unusual. More than half of the 20 known FU\,Ori objects drive collimated jets, and almost all have been shown to possess powerful winds \citep{reipurth:2010}. The energy fluxes of the outflows from FU\,Ori objects far exceed the luminosities of the pre-outburst stars \citep{calvet:1998}. The outflows must therefore be powered by accretion.  This result is consistent with the strong correlation between measured mass-loss rates and accretion rates \citep{hartmann:2009}.

FU\,Ori outbursts are believed to be connected to periods of increased accretion \citep[in which the accretion rate may reach as high as 10$^{-4}$ M$_{\sun}$ yr$^{-1}$;][]{hartmann:1985,hartmann:1996}. An FU\,Ori object may experience hundreds of outbursts \citep{covey:2011,miller:2011}. During these outbursts the central object may gain as much as 10\% to 25\% of its final mass \citep{herbig:1977,hartmann:1996,greene:2008}. The time-scale between FU\,Ori outbursts is typically of the order of 10$^3$ years \citep{hartmann:2009}. There are, however, subtle differences between the observed light curves of some objects \citep[see][for the details]{reipurth:2010}. A number of mechanisms may therefore be at work. These include: a pile-up of infalling matter in the disk which, when accreted, results in a short-lived outburst \citep[e.g.,][]{hartmann:1996,zhu:2009}; disk thermal instabilities \citep{bell:1994,armitage:2001}; disk fragmentation processes \citep{vorobyov:2006,vorobyov:2010}; and disk perturbations caused by a planet \citep{clarke:2005}, a close encounter with a companion in an eccentric orbit \citep{bonnell:1992,artymowicz:1996}, or an encounter with another member of a dense star cluster \citep{pfalzner:2008}. Distinguishing between these models is observationally very difficult. Even so, based on the symmetry often observed in HH jets and MHO outflows (HH\,212 is the most striking example, though this jet is not known to be driven by an FU\,Ori object), one may at least make a crude estimate of the time-scale between outburst events.

It is not unreasonable to assume that the FU\,Ori objects found in the Braid Nebula region, V2494\,Cyg and V2495\,Cyg, have undergone several outbursts in the last $\sim$10$^5$ years. The HH objects and MHOs detected in the bipolar flow lobes associated with each young star probably represent a fossil record of these outburst events. V2494\,Cyg drives an outflow, traced by HH\,381/966/967 and MHO\,900-904 (see e.g. Fig.\,\ref{fig:v2494cygwide} and Table\,\ref{tab:outflows}) that extends over $\sim$10 pc; the V2495\,Cyg outflow extends over almost 3 pc when measured from end-to-end and likewise includes a number of HH objects (HH\,629) and MHOs (MHO\,916/924/927: Table\,\ref{tab:outflows}). Features seen in the opposing lobes of each bipolar outflow that are at equal distances from the driving source may be associated with the same outburst event. For example, MHO~900 and 904 are roughly equidistant from the driving source, V2494\,Cyg. The propagation speeds of these shock fronts have not been measured, though MHOs typically possess proper motions of $\sim$10$^2$ km\,s$^{-1}$ \citep[e.g.][]{davis:2009}, while similar tangential velocities have been measured for HH objects \citep[e.g.][]{eisloffel:1994}. The dynamical ages of MHO~900 and 904 (i.e. the time for each shock front to propagate over the observed distance from V2494\,Cyg) are of the order of $10^5$ years. This means that V2494\,Cyg has been outbursting for at least 100,000 years.

MHO 902 and HH 966 may be associated with a more recent outburst event from V2494\,Cyg; with an average distance of 2.7\,pc, their dynamical ages are of the order of $10^4$ years. Closer to the source, MHO\,901 has an estimated dynamical age of 3.5$\times$10$^3$\,yrs (although no corresponding feature is detected in the opposing outflow lobe). The HH\,381\,jet is the shock feature closest to the central source \citep{magakian:2010}. Assuming the same propagation speed, the jet length indicates a dynamical time scale of $\sim$75 years. Since V2494\,Cyg is currently undergoing an outburst, the time scale between (recent) outbursts may be as short as $\sim$100 years.

Similar analysis is possible for the Braid star, V2495\,Cyg (the most distant knot, MHO\,927B/HH\,635, corresponds to an age of 3$\times$10$^4$ yrs, MHO\,924 and HH\,635 knots F and G correspond to 7.5$\times$10$^3$ years, HH\,635E to 4.5$\times$10$^3$ years, and HH\,635D to 2.3$\times$10$^3$ years), although no HH objects/compact jet or MHOs are observed very close to the driving source. We stress, however, that the ambient medium may affect the ``flow lengths" measured here. The propagation distances are not corrected for inclination angle, and not all shock working surfaces may be visible (some features may be too faint and/or may be hidden by extinction). At best, these estimates are probably only accurate to within an order of magnitude.

\subsection{Star formation in the region: An Updated picture}

The star formation activity within the studied region (Fig.\,\ref{fig:field-overview}) has been proven numerous times in our previous works \citep{movsessian:2003,movsessian:2006,aspin:2009,aspin:2011,magakian:2010}. In this current study, which focuses on the near-infrared outflows and their driving sources, we add another layer to the existing information on the region. These new data undoubtedly lead to a broader picture of star formation activity occurring in the Braid Nebula region.

Careful examination of the surveyed region using the new results discussed in Sections\,\ref{sec:outflparam} and \ref{sec:protcont} show a clear association between the outflows, their driving sources and the main regions of high extinction, namely LDN\,1003 and 1004 (cf. Fig.\,\ref{fig:field-overview}). These high-$A_V$ regions extend roughly North-West to South-East and broadly speaking coincide with HiRes 100$\mu$m emission features and CSO 1.1mm peaks, as can be seen in Fig.\,\ref{fig:outflows}. There is a general correlation between the CSO 1.1mm emission and A$_V$ data \citep{aspin:2009}, the A$_V$ data being more sensitive to the low extinct areas, whereas the CSO 1.1mm emission traces the very high extinction areas \citep{kirk:2006,davis:2008}. Somewhat surprisingly there is a large 1.1~mm dust core ~10-20\arcmin\ north of LDN~1004 that does not coincide with a region of high extinction or any young stars or outflow features (HH objects or MHOs). This peak may be too diffuse to produce star forming cores; note also that the $A_V$ map is higher-resolution than the CSO data.

There seems to be a distinct difference between the LDN\,1003 and 1004 regions in general. The A$_V$ map in Fig.\,\ref{fig:outflows} suggests that LDN\,1004 has more diffuse and extended structure, while LDN\,1003 is more filamentary. There is less extended IRAS HiRes 100$\mu$m emission in LDN\,1003 than in the East (LDN\,1004) and north-east, where there are isolated patches of IRAS HiRes 100$\mu$m emission. This indicates that the area could still be gravitationally unbound or located further away.

While most of the outflows (23 out of 28) are being driven from protostars in LDN\,1003, we were able to detect only five outflows emanating from sources located in LDN\,1004. The same stark difference is evident when one considers the protostellar content, where most of the identified protostars are located in the LDN\,1003 cloud (Fig.\,\ref{fig:outflows}). 

The differences outlined above suggest that LDN\,1003 is going through an active phase of star formation. By comparison, star formation may be just starting in LDN\,1004, as evidenced by the low number of outflows and protostars detected there. The extended diffuse gas traced by IRAS HiRes 100$\mu$m emission could eventually form dense cores from which a new generation of protostars could form. Based on our suite of observations of the region, we are not able to state conclusively why star formation started in LDN\,1003 earlier than in LDN\,1004. All of the evidence seems to suggest that star formation is moving from LDN\,1003 toward LDN\,1004.

Finally, we caution that the observed differences between LDN\,1003 and LDN\,1004 could simply be a result of LDN\,1004 being located further away than LDN\,1003. For this to explain the apparent paucity of star formation in LDN\,1004, it would need to be at least twice as far away as LDN\,1003. There is no evidence for this in the literature.

\section{SUMMARY}
\label{sec:summary}

We have studied a one-square-degree field centred on the Braid Nebula using the near-infrared H$_2$ 1-0 S(1) line to trace the outflows and shocks driven by the many young stars found in the region \citep{aspin:2009}. Our analysis is complemented by optical H$\alpha$ and [SII] observations of HH shocks \citet{magakian:2010} as well as archival multi-waveband photometry datasets. Our main findings and conclusion are as follows:

\begin{enumerate}
	\item We have detected 37 complex MHOs comprising 79 individual knots and bow shocks; 32 of the 79 H$_2$ features are associated with optical HH objects. This significant increase in the number of detected outflow features is expected due to the considerable optical extinction present in the area.  
	\item The newly-detected MHOs and previously discovered HH objects enable us to discern 20 outflows from which 15 are bipolar and five are monopolar. We have also detected a further 8 possible outflows of which two are bipolar and six are monopolar.
	\item We were also able to identify the driving sources of many of the outflows with high certainty using positional and colour-colour analysis. Of the 18 identified sources two drive multiple, almost perpendicular outflows, suggesting a multiple system of young stars in each case. All but a handful of sources have the distinct colours of embedded protostars prescribed to the Class\,I evolutionary phase. Three of the sources are undetected at near- and mid-infrared wavelengths, suggesting a Class\,0 protostellar phase. We also suggest that the protostellar population present in this star formation region differs slightly from that of Taurus-Auriga, in the sense that our objects are possibly more energetic and/or more massive.
	\item Both FU\,Ori outburst type stars drive significant bipolar outflows. Our analysis suggest a connection between outflow ejection and repetitive FU\,Ori outburst events.
	\item Consideration of the multi-waveband photometry, combined with the outflow information described in detail in this paper, suggests that the star-formation in the Braid Nebula Star Formation region is progressing from LDN\,1003 in the West/South-West to LDN\,1004 in the East/South-East.
\end{enumerate}

\begin{acknowledgements}
This research has made use of the SIMBAD database, operated at CDS, Strasbourg, France as well as software provided by the UK's AstroGrid Virtual Observatory Project, which is funded by the Science and Technology Facilities Council and through the EU's Framework 6 programme. This research has made use of the NASA/ IPAC Infrared Science Archive, which is operated by the Jet Propulsion Laboratory, California Institute of Technology, under contract with the National Aeronautics and Space Administration. In particular data from IRAS, MSX, AKARI and 2MASS missions were used. \emph{The authors wish to recognise and acknowledge the very significant cultural role and reverence that the summit of Mauna Kea has always had within the indigenous Hawaiian community. We are most fortunate to have the opportunity to conduct observations from this sacred mountain.}

\end{acknowledgements}

\bibliographystyle{aa}
\bibliography{main-library}


\onecolumn

\begin{appendix}
\section{Individual Molecular Hydrogen emission-line Objects}
\label{sec:indmho}

Our analysis of the entire one-square-degree field (Fig.\,\ref{fig:field-overview}) has revealed more than 200 individual H$_2$ 1-0 S(1) emission line knots, bow shocks, and spikes which we group into 37 MHOs according to their positions and overall morphology. The MHO numbers from 900 to 936 were assigned according to the rising RA of each feature. In Table\,\ref{tab:mhos} we list the names and coordinates of the detected MHOs in the first, second and third columns, respectively. Columns four and five list the observed and de-reddened fluxes for each MHO. Column six records the visual extinction (A$_V$) obtained from the extinction map of the region presented in Fig.\,\ref{fig:outflows}. Column seven and eight describe the morphology and list any Herbig-Haro (HH) objects associated with the 37 individual MHOs.

\begin{longtable}{lcccccll}
	\caption{MHOs detected and identified in our survey\label{tab:mhos}}\\
	\hline\hline
	MHO &  RA & Dec & \multicolumn{2}{c}{Flux(W/m$^2$)$\times$10$^{-17}$} & A$_V$ &Morphology & HH Assoc.\\
		\#  &  (2000) & (2000) & obs. & dered. & & & \\
	\hline
	\endfirsthead
	\caption{continued.}\\
	\hline\hline
	MHO &  RA & Dec & \multicolumn{2}{c}{Flux(W/m$^2$)$\times$10$^{-17}$} & A$_V$ & Morphology & HH Assoc.\\
		\#  &  (2000) & (2000) & obs. & dered. & & & \\
	\hline
	\endhead
	\hline
	\endfoot
	900  & 20:58:07.7 & +52:48:59 & 51.28 & 51.92 & 0.12 & irregular              & \ldots    \\  
	901A & 20:58:21.7 & +52:27:33 & 2.07  & 2.66  & 2.43 & tip of the bow         & HH 382E    \\ 
	901B & 20:58:21.9 & +52:27:39 & 3.70  & 4.75  & 2.43 & wings of the bow       & HH 382E    \\ 
	901C & 20:58:22.7 & +52:27:48 & 2.07  & 2.68  & 2.50 & compact                & \ldots    \\  
	901D & 20:58:22.4 & +52:27:54 & 4.75  & 6.15  & 2.51 & compact                & HH 382F    \\ 
	902A & 20:58:24.5 & +52:17:18 & 14.85 & 15.10 & 0.16 & bright \& compact      & HH 382     \\ 
	902B & 20:58:25.2 & +52:18:10 & 4.20  & 4.27  & 0.16 & faint knot             & \ldots    \\  
	902C & 20:58:26.4 & +52:18:29 & 44.32 & 45.24 & 0.20 & bright bow-like        & \ldots    \\  
	902D & 20:58:26.9 & +52:17:59 & 4.64  & 4.75  & 0.23 & compact                & HH 382D    \\ 
	903  & 20:58:45.5 & +52:22:48 & 38.04 & 51.94 & 3.02 & diffuse pair           & \ldots    \\  
	904A & 20:58:46.3 & +52:05:34 & 3.05  & 3.17  & 0.38 & faint \& diffuse       & \ldots    \\  
	904B & 20:58:49.9 & +52:06:17 & 2.95  & 3.08  & 0.41 & faint \& diffuse       & \ldots    \\  
	905  & 20:58:56.1 & +52:30:41 & 3.04  & 5.29  & 5.37 & compact, bow-like      & \ldots    \\  
	906A & 20:58:55.1 & +52:34:59 & 4.41  & 6.01  & 3.00 & faint \& diffuse       & HH 380F    \\ 
	906B & 20:59:02.6 & +52:35:45 & 21.66 & 26.35 & 1.90 & diffuse                & HH 380B(?) \\ 
	906C & 20:59:08.3 & +52:36:35 & 10.82 & 13.49 & 2.14 & extended, bow-like     & HH 380C    \\ 
	907A & 20:59:05.1 & +52:29:18 & 1.53  & 2.83  & 5.95 & faint \& compact       & \ldots    \\  
	907B & 20:59:06.7 & +52:29:52 & 73.23 & 127.03& 5.34 & complex bow shock      & HH 970     \\ 
	908A & 20:59:05.9 & +52:22:32 & 15.44 & 21.63 & 3.27 & right, compact bow     & \ldots    \\  
	908B & 20:59:07.8 & +52:22:30 & 157.98& 217.51& 3.10 & right \& compact       & \ldots    \\  
	908C & 20:59:08.6 & +52:22:27 & 7.77  & 10.47 & 2.89 & compact pair of bows   & \ldots    \\  
	908D & 20:59:11.0 & +52:22:29 & 4.87  & 6.42  & 2.68 & compact pair of knots  & HH 974B(?) \\ 
	909A & 20:59:10.5 & +52:33:04 & 2.91  & 3.60  & 2.07 & faint shock front      & \ldots    \\  
	909B & 20:59:09.5 & +52:33:27 & 5.28  & 6.22  & 1.59 & compact knot           & \ldots    \\  
	909C & 20:59:08.8 & +52:33:39 & 2.51  & 3.07  & 1.94 & broken bow             & \ldots    \\  
	910A & 20:59:11.9 & +52:34:12 & 129.79& 143.45& 0.97 & complex bow shock      & HH 380(A)  \\ 
	910B & 20:59:15.2 & +52:34:19 & 5.74  & 6.24  & 0.81 & faint, diffuse flow    & \ldots    \\  
	910C & 20:59:14.9 & +52:35:10 & 2.67  & 2.98  & 1.05 & faint bow-like         & HH 380E    \\ 
	911  & 20:59:46.3 & +52:33:09 & 80.50 & 141.53& 5.47 & very bright bow        & HH 627     \\ 
	912A & 20:59:51.3 & +52:16:29 & 154.55& 200.43& 2.52 & complex of bow shocks  & \ldots    \\  
	912B & 20:59:56.7 & +52:17:13 & 7.04  & 9.51  & 2.92 & bright knot            & \ldots    \\  
	913A & 20:59:50.3 & +52:40:38 & 11.99 & 16.10 & 2.86 & bright \& compact      & HH 675     \\ 
	913B & 20:59:51.3 & +52:40:34 & 8.38  & 11.26 & 2.86 & broken bow             & \ldots    \\  
	913C & 20:59:53.5 & +52:39:59 & 5.97  & 8.31  & 3.21 & diffuse, bow-like      & \ldots    \\  
	914A & 21:00:08.6 & +52:25:55 & 41.61 & 76.08 & 5.85 & bright \& complex      & \ldots    \\  
	914B & 21:00:08.5 & +52:26:07 & 7.29  & 13.71 & 6.12 & diffuse                & \ldots    \\  
	915A & 21:00:03.8 & +52:17:28 & 5.63  & 7.61  & 2.92 & shock front            & \ldots    \\  
	915B & 21:00:06.9 & +52:18:01 & 5.39  & 7.37  & 3.03 & chain of faint bows    & \ldots    \\  
	915C & 21:00:11.7 & +52:18:25 & 16.97 & 26.31 & 4.25 & chain of knots         & \ldots    \\  
	915D & 21:00:15.4 & +52:19:02 & 3.20  & 5.42  & 5.10 & faint knot             & \ldots    \\  
	916  & 21:00:10.9 & +52:29:01 & 7.78  & 15.85 & 6.90 & pair of faint knots    & HH 629(?)  \\ 
	917  & 21:00:12.1 & +52:27:15 & 8.25  & 16.42 & 6.67 & compact bow shock      & HH 670     \\ 
	918  & 21:00:12.1 & +52:22:42 & 4.69  & 8.48  & 5.74 & faint \& diffuse       & \ldots    \\  
	919A & 21:00:18.2 & +52:28:07 & 7.29  & 15.24 & 7.15 & bright \& compact      & HH 631E    \\ 
	919B & 21:00:19.2 & +52:27:53 & 3.17  & 6.61  & 7.12 & compact                & HH 631D    \\ 
	920A & 21:00:18.5 & +52:25:55 & 24.59 & 46.23 & 6.12 & complex bow shock      & HH 630A    \\ 
	920B & 21:00:16.1 & +52:26:04 & 4.10  & 7.71  & 6.12 & faint \& diffuse       & \ldots    \\  
	921A & 21:00:22.8 & +52:26:22 & 24.59 & 48.63 & 6.61 & complex bow shock      & HH 977     \\ 
	921B & 21:00:23.1 & +52:26:46 & 3.00  & 6.02  & 6.75 & faint shock            & \ldots    \\  
	921C & 21:00:20.7 & +52:26:41 & 1.50  & 2.93  & 6.48 & faint \& diffuse       & \ldots    \\  
	922A & 21:00:28.9 & +52:20:25 & 5.93  & 10.00 & 5.07 & faint, extended bow    & \ldots    \\  
	922B & 21:00:32.4 & +52:21:02 & 30.72 & 51.67 & 5.04 & broken bow shock       & \ldots    \\  
	922C & 21:00:38.3 & +52:21:28 & 139.44& 227.37& 4.74 & pair of bright bows    & \ldots    \\  
	923A & 21:00:30.5 & +52:29:26 & 13.84 & 24.56 & 5.56 & compact bow shock      & HH 448A    \\ 
	923B & 21:00:31.7 & +52:29:18 & 50.61 & 92.16 & 5.81 & group of shocks        & HH 448B,D  \\ 
	923C & 21:00:36.2 & +52:29:36 & 42.63 & 67.67 & 4.48 & complex bow shock      & HH 448G    \\ 
	923D & 21:00:37.6 & +52:29:24 & 79.95 & 123.81& 4.24 & complex bow shocks     & HH 448I    \\ 
	924  & 21:00:42.5 & +52:31:43 & 30.15 & 41.77 & 3.16 & bright, compact bow    & HH 635A    \\ 
	925  & 21:00:42.7 & +52:25:11 & 9.24  & 15.75 & 5.17 & bright \& compact      & HH 633     \\ 
	926A & 21:00:37.7 & +52:28:25 & 1.49  & 2.52  & 5.11 & compact \& faint       & HH 634A    \\ 
	926B & 21:00:41.4 & +52:27:43 & 57.83 & 95.57 & 4.87 & complex bow shock      & HH 634B    \\ 
	926C & 21:00:48.7 & +52:26:49 & 5.68  & 8.79  & 4.23 & bright shock front     & HH 634C    \\ 
	927A & 21:00:52.0 & +52:33:27 & 10.00 & 13.18 & 2.68 & chain of knots         & HH 635B    \\ 
	927B & 21:00:56.0 & +52:33:50 & 5.21  & 6.48  & 2.11 & faint bow shock        & HH 635C    \\ 
	928  & 21:01:58.8 & +52:29:37 & 3.80  & 6.50  & 5.20 & faint pair of shocks   & \ldots    \\  
	929A & 21:01:57.1 & +52:29:00 & 5.28  & 10.50 & 6.66 & faint \& diffuse       & \ldots    \\  
	929B & 21:01:59.3 & +52:28:56 & 5.91  & 11.69 & 6.61 & faint knot             & \ldots    \\  
	930A & 21:02:02.3 & +52:28:44 & 11.88 & 24.58 & 7.05 & faint shock            & \ldots    \\  
	930B & 21:02:02.3 & +52:28:38 & 4.68  & 9.68  & 7.05 & faint \& diffuse       & \ldots    \\  
	930C & 21:02:00.9 & +52:28:08 & 5.83  & 12.89 & 7.69 & faint \& diffuse       & \ldots    \\  
	931  & 21:02:02.8 & +52:19:24 & 13.86 & 30.48 & 7.64 & faint, bow like shock  & \ldots    \\  
	932A & 21:02:03.5 & +52:20:07 & 30.35 & 69.70 & 8.06 & complex group of knots & \ldots    \\  
	932B & 21:02:02.6 & +52:20:06 & 25.09 & 59.49 & 8.37 & diffuse \& extended    & \ldots    \\  
	932C & 21:02:02.2 & +52:19:53 & 4.87  & 11.35 & 8.20 & faint knot             & \ldots    \\  
	933A & 21:02:09.9 & +52:28:17 & 9.38  & 19.69 & 7.19 & bright shock           & \ldots    \\  
	933B & 21:02:17.7 & +52:27:23 & 4.30  & 10.24 & 8.41 & faint, bow-like        & \ldots    \\  
	934  & 21:02:10.7 & +52:35:57 & 13.95 & 15.23 & 0.85 & pair of bow shocks     & HH 978     \\ 
	935A & 21:02:08.3 & +52:29:32 & 2.12  & 3.57  & 5.04 & faint \& diffuse       & \ldots    \\  
	935B & 21:02:11.8 & +52:30:38 & 2.10  & 3.00  & 3.47 & faint bow shock        & \ldots    \\  
	936  & 21:02:21.7 & +52:26:09 & 28.46 & 65.84 & 8.13 & bright group of knots  & \ldots    \\
\end{longtable}

\end{appendix}

\begin{appendix}
\section{Multi-waveband data of the sources}
\label{sec:multisource}

The driving sources of the MHOs listed in Table\,\ref{tab:mhos} are presented in Table\,\ref{tab:sources}.  Here we list the appropriate IDs from the 2MASS, AKARI/IRC, MSX6C, AKARI/FIS, IRAS and CSO 1.1mm \citep{aspin:2011} catalogues. Table\,\ref{tab:photom} lists the near to mid-infrared photometric magnitudes and colour indices for the same set of outflow sources (see also Figs.\,\ref{fig:ccdjhhk} and \ref{fig:ccdmir}).

\begin{table}
	\caption{Multi-waveband Source Associations\label{tab:sources}}
	\scriptsize
	\centering
	\begin{tabular}{lcccccc}
		\hline\hline
Source &  2MASS & AKARI/IRC & MSX6C & AKARI/FIS & IRAS\tablefootmark{a} & CSO 1.1mm\\
Name	& ID & ID & ID & ID & ID & ID \\
		\hline\\
Z20566+5213
	& \ldots
	& \ldots
	& \ldots
	& \ldots
	& Z20566+5213 
	& \ldots \\
V2494Cyg
	& 20582109+5229277 
	& 2058211+522927 
	& G091.6383+04.4007
	& 2058198+522930 
	& I20568+5217
	& G091.638+04.398 (\#37)\\
I20573+5221
	& 20585009+5232543
	& 2058500+523255
	& G091.7301+04.3831
	& 2058498+523312
	& I20573+5221
	& G091.741+04.377 (\#47)\\
CN5
	& 20590419+5221448
	& \ldots
	& \ldots
	& \ldots
	& \ldots
	& \ldots\\
core \#44
	& \ldots
	& \ldots
	& \ldots
	& 2059087+523221
	& \ldots
	& G091.742+04.338 (\#44)\\
I20575+5210
	& \ldots
	& \ldots
	& \ldots
	& 2059066+522228
	& I20575+5210
	& G091.626+04.234 (\#18)\\
CN4
	& 20590895+5222392
	& \ldots
	& \ldots
	& \ldots
	& \ldots
	& \ldots\\
CN6N
	& 20594000+5234361 
	& 2059400+523436
	& \ldots
	& \ldots
	& R20581+5222
	& \ldots\\
CN6(core \#51)
	& 20594071+5234135
	& 2059408+523414
	& G091.8313+04.2995
	& 2059430+523355
	& R20581+5222
	& G091.838+04.341 (\#51)\\
R20582+5221
	& 20594619+5233216
	& 2059462+523322
	& G091.8296+04.2799
	& \ldots
	& R20582+5221
	& \ldots\\
I20583+5228
	& 20595184+5240205
	& 2059518+524020
	& G091.9269+04.3452
	& 2059513+524021
	& I20583+5228
	& G091.926+04.343 (\#54)\\
CN3N
	& 21000376+5234290
	& 2100038+523430
	& G091.8730+04.2583
	& \ldots
	& I20585+5222
	& \ldots \\
Cyg 19
	& 21001159+5218172
	& \ldots
	& \ldots
	& 2100124+521831
	& \ldots
	& G091.689+04.064 (\#7)\\
CN2
	& 21001656+5226230
	& \ldots 
	& \ldots
	& \ldots 
	& \ldots 
	& \ldots\\
CN8
	& 21001903+5227281
	& 2100191+522728
	& \ldots
	& \ldots
	& \ldots
	& \ldots\\
IRAS 15N
	& 21002140+5227094
	& 2100215+522710
	& \ldots
	& 2100217+522712
	& I20588+5215
	& G091.808+04.142 (\#30)\\
I20588+5221
	& 21002143+5222570
	& 2100215+522258
	& G091.7575+04.0988
	& \ldots
	& I20588+5221
	& G091.759+04.106 (\#20)\\
V2495 Cyg
	& \ldots
	& 2100253+523017
	& \ldots
	& 2100249+523018
	& \ldots
	& G091.861+04.172 (\#39)\\
CN1
	& 21003517+5233244
	& 2100352+523324
	& \ldots
	& \ldots
	& I20590+5221
	& \ldots\\
core \#33
	& \ldots
	& 2100388+522757
	& \ldots
	& 2100387+522753
	& \ldots
	& G091.855+04.119 (\#33)\\
IRAS 14
	& 21004237+5226007
	& 2100423+522600
	& \ldots
	& 2100422+522552
	& I20591+5214
	& G091.835+04.087 (\#24)\\
21015977+5220127
	& 21015977+5220127
	& 2101598+522013
	& G091.8861+03.8801
	& \ldots
	& \ldots
	& \ldots\\
I21005+5217
	& 21020547+5228544
	& 2102055+522854
	& G092.0055+03.9642
	& 2102042+522858
	& I21005+5217
	& \ldots \\
HD 200594
	& 21022391+5234328
	& \ldots
	& \ldots
	& \ldots
	& \ldots
	& \ldots\\
	\hline
\end{tabular}
	\tablefoot{This table contains the multi-waveband source identifications for all outflow driving sources discussed in the text.
	\tablefoottext{a}{The letters in front of IRAS sources denote: 
                         \textbf{I} $=$ IRAS PSC Version 2.0 (IPAC 1986), 
                         \textbf{R} $=$ IRAS PS Reject Catalogue \citet{infrared-astronomical-satellite-iras-catalogs:2007} and 
                         \textbf{Z} $=$ IRAS Faint Source Reject Catalogue \citep{moshir:2008}.}
	}
\end{table}

\begin{table}
	\caption{Near to Mid-Infrared Photometry}\label{tab:photom}
	\scriptsize
	\centering
	\begin{tabular}{lcccccccccc}
		\hline\hline
Source	& J & H & K & S9W\tablefootmark{a} & L18W\tablefootmark{a} & J - H & H - K & J - K
		& K - S9W & K - L18W\\
Name & [mag] & [mag] & [mag] & [mag] & [mag] & [mag] & [mag] & [mag] & [mag] & [mag]\\
\hline\\
	Z20566+5213
		& \textbf{17.30 $\pm$0.03}
		& \textbf{15.94 $\pm$0.03}
		& \textbf{15.14 $\pm$0.03}
		& \ldots 
		& \ldots 
		& 1.37$\pm$0.04
		& 0.79$\pm$0.04
		& 2.16$\pm$0.04
		& \ldots
		& \ldots\\
	V2494Cyg
		& 11.54$\pm$0.03
		& 9.81$\pm$0.03
		& 8.31$\pm$0.02
		& 2.02$\pm$0.50
		& -0.72$\pm$0.54
		& 1.73$\pm$0.04
		& 1.51$\pm$0.03
		& 3.24$\pm$0.04
		& 6.28$\pm$0.50
		& 9.63$\pm$0.60\\
	I20573+5221
		& 13.62$\pm$0.03
		& 11.30$\pm$0.02
		& 9.73$\pm$0.02
		& 6.22$\pm$0.03
		& 3.37$\pm$0.08
		& 2.32$\pm$0.03
		& 1.57$\pm$0.03
		& 3.89$\pm$0.03
		& 3.51$\pm$0.04
		& 6.37$\pm$0.08\\
	CN5
		& 14.49$\pm$0.04
		& 13.08$\pm$0.03
		& 12.19$\pm$0.03
		& \ldots
		& \ldots
		& 1.42$\pm$0.04
		& 0.89$\pm$0.04
		& 2.30$\pm$0.05
		& \ldots
		& \ldots\\
	core \#44
		& \ldots
		& \ldots
		& \ldots
		& \ldots
		& \ldots
		& \ldots
		& \ldots
		& \ldots
		& \ldots
		& \ldots\\
	I20575+5210
		& \ldots
		& \ldots
		& \ldots
		& \ldots
		& \ldots
		& \ldots
		& \ldots
		& \ldots
		& \ldots
		& \ldots\\
	CN4
		& 13.28$\pm$0.03
		& 12.05$\pm$0.02
		& 11.32$\pm$0.03
		& \ldots
		& \ldots
		& 1.23$\pm$0.03
		& 0.73$\pm$0.03
		& 1.96$\pm$0.04
		& \ldots
		& \ldots\\
	CN6N
		& 14.57$\pm$0.03
		& 10.61$\pm$0.02
		& 8.66$\pm$0.02
		& 7.04$\pm$0.01
		& \ldots
		& 3.95$\pm$0.04
		& 1.95$\pm$0.03
		& 5.90$\pm$0.04
		& 1.63$\pm$0.02
		& \ldots\\
	CN6(core \#51)
		& 15.21$\pm$0.06
		& 13.03$\pm$0.03
		& 11.32$\pm$0.03
		& 5.99$\pm$0.02
		& 3.66$\pm$0.04
		& 2.18$\pm$0.07
		& 1.71$\pm$0.04
		& 3.89$\pm$0.07
		& 5.33$\pm$0.04
		& 7.66$\pm$0.05\\
	R20582+5221
		& 13.34$\pm$0.02
		& 11.94$\pm$0.02
		& 11.07$\pm$0.02
		& 6.58$\pm$0.02
		& 3.86$\pm$0.04
		& 1.40$\pm$0.03
		& 0.87$\pm$0.03
		& 2.27$\pm$0.03
		& 4.49$\pm$0.03
		& 7.23$\pm$0.05\\
	I20583+5228
		& 14.48$\pm$0.04
		& 12.29$\pm$0.02
		& 10.97$\pm$0.02
		& 6.17$\pm$0.03
		& 3.24$\pm$0.07
		& 2.19$\pm$0.04
		& 1.34$\pm$0.03
		& 3.52$\pm$0.04
		& 4.80$\pm$0.03
		& 7.73$\pm$0.07\\
	CN3N
		& 13.32$\pm$0.03
		& 11.26$\pm$0.02
		& 9.99$\pm$0.02
		& 6.03$\pm$0.02
		& 4.10$\pm$0.07
		& 2.06$\pm$0.03
		& 1.27$\pm$0.03
		& 3.33$\pm$0.03
		& 3.96$\pm$0.03
		& 5.89$\pm$0.08\\
	Cyg 19
		& 11.38$\pm$0.02
		& 10.48$\pm$0.02
		& 10.02$\pm$0.02
		& \ldots
		& \ldots
		& 0.91$\pm$0.03
		& 0.46$\pm$0.02
		& 1.37$\pm$0.03
		& \ldots
		& \ldots\\
	CN2
		& 12.60$\pm$0.02
		& 11.47$\pm$0.02
		& 10.67$\pm$0.02
		& \ldots
		& \ldots
		& 1.13$\pm$0.03
		& 0.80$\pm$0.03
		& 1.93$\pm$0.03
		& \ldots
		& \ldots\\
	CN8
		& 11.69$\pm$0.02
		& 10.63$\pm$0.02
		& 9.81$\pm$0.02
		& 5.85$\pm$0.05
		& 3.38$\pm$0.12
		& 1.06$\pm$0.03
		& 0.82$\pm$0.03
		& 1.88$\pm$0.03
		& 3.96$\pm$0.05
		& 6.43$\pm$0.12\\
	IRAS 15N
		& 11.15$\pm$0.03
		& 10.23$\pm$0.03
		& 9.40$\pm$0.04
		& 4.02$\pm$0.11
		& 1.16$\pm$1.38
		& 0.94$\pm$0.04
		& 0.82$\pm$0.05
		& 1.75$\pm$0.05
		& 5.38$\pm$0.11
		& 8.24$\pm$1.38\\
	I20588+5221
		& \textbf{20.96$\pm$0.50}
		& 16.92$\pm$0.05
		& 14.03$\pm$0.04
		& 6.10$\pm$0.01
		& 3.83$\pm$0.05
		& 4.04$\pm$0.50
		& 2.89$\pm$0.07
		& 6.93$\pm$0.50
		& 7.93$\pm$0.04
		& 10.20$\pm$0.07\\
	V2495Cyg
		& 13.57$\pm$0.02
		& 11.48$\pm$0.02
		& 10.29$\pm$0.02
		& 4.66$\pm$0.03
		& 1.64$\pm$0.15
		& 2.09$\pm$0.03
		& 1.20$\pm$0.03
		& 3.28$\pm$0.03
		& 5.63$\pm$0.03
		& 8.65$\pm$0.15\\
	CN1
		& 15.27$\pm$0.06
		& 13.86$\pm$0.04
		& 12.52$\pm$0.03
		& \ldots
		& 4.22$\pm$0.08
		& 1.41$\pm$0.07
		& 1.34$\pm$0.05
		& 2.75$\pm$0.07
		& \ldots
		& 8.30$\pm$0.08\\
	core \#33
		& \textbf{21.94$\pm$0.50}
		& \textbf{21.21$\pm$0.50}
		& 15.25$\pm$0.02
		& 7.01$\pm$0.01
		& 4.34$\pm$0.04
		& 0.73$\pm$0.71
		& 5.95$\pm$0.50
		& 6.69$\pm$0.50
		& 8.24$\pm$0.02
		& 10.91$\pm$0.04\\
	IRAS 14
		& \textbf{20.43$\pm$0.50}
		& 16.91$\pm$0.02
		& 14.22$\pm$0.09
		& \ldots
		& 4.62$\pm$0.12
		& 3.52$\pm$0.50
		& 2.69$\pm$0.09
		& 6.21$\pm$0.51
		& \ldots
		& 9.60$\pm$0.15\\
	21015977+5220127
		& 12.05$\pm$0.02
		& 9.17$\pm$0.02
		& 7.79$\pm$0.02
		& 6.57$\pm$0.01
		& \ldots
		& 2.88$\pm$0.03
		& 1.37$\pm$0.03
		& 4.26$\pm$0.03
		& 1.22$\pm$0.03
		& \ldots\\
	I21005+5217
		& 13.46$\pm$0.02
		& 11.70$\pm$0.02
		& 10.52$\pm$0.02
		& 4.73$\pm$0.06
		& 1.77$\pm$0.41
		& 1.76$\pm$0.03
		& 1.18$\pm$0.03
		& 2.94$\pm$0.03
		& 5.79$\pm$0.06
		& 8.75$\pm$0.41\\
	HD 200594
		& 8.67$\pm$0.03
		& 8.70$\pm$0.02
		& 8.69$\pm$0.02
		& \ldots
		& \ldots
		& -0.03$\pm$0.04
		& 0.01$\pm$0.03
		& -0.03$\pm$0.03
		& \ldots
		& \ldots\\
		\hline
	\end{tabular}
	\tablefoot{Near to mid-infrared magnitudes and colour-colour differences used in Figs.\,\ref{fig:ccdjhhk} and \ref{fig:ccdmir} for the sources listed in  Table\,\ref{tab:sources}. In cases (highlighted in bold text) when the 2MASS PSC has no identifiable source we used our WFCAM J, H and K band data to calculate the magnitudes and/or magnitude limits at 1$\sigma$ level.
	\tablefoottext{a}{Mid-infrared AKARI/IRC 9 and 18$\mu$m fluxes from all-sky survey catalogue \citep{ishihara:2010} converted into magnitudes.}
	}
\end{table}

\end{appendix}

\begin{appendix}

\twocolumn

\section{Individual Outflow Regions}
\label{sec:outfldet}
\begin{figure*}
	\centering
		\includegraphics[width=16cm]{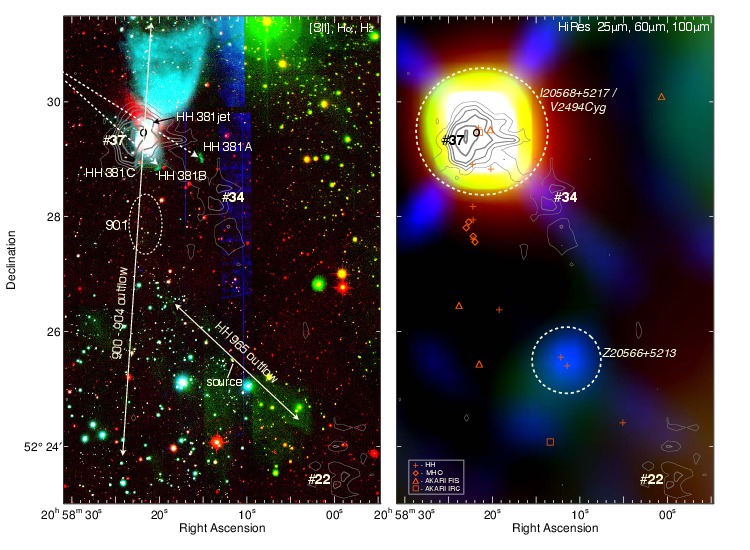}
	\caption{Colour-composite images of the area that includes IRAS\,20568+5217, V2494\,Cyg and numerous other related features. Contours show the CSO 1.1mm data with values 40, 60, 80, 100, 140 and 180 mJy. The hashed numbers are the designated CSO cores from \citet{aspin:2011}. The \textbf{left panel} is a composite of the optical [SII] (blue), H$\alpha$ (green) and near-infrared H$_2$ 1-0 S(1) (red) line data, where the incomplete coverage in [SII] causes the vertical stripe artefact. The solid-line and dashed-line arrows show the propagation direction of the confirmed and questionable outflows, respectively. The dashed line ellipse in this case indicates the extent of the MHO\,901 sub-knots. The \textbf{right panel} shows the same area constructed from IRAS HiRes 25$\mu$m (blue), 60$\mu$m (green) and 100$\mu$m (red) bands, where the positions of known HH objects, MHOs and sources from the AKARI survey (Tab.\,\ref{tab:sources}) are marked and explained in the legend. The dashed-line circles indicate the apparent extend of the sources discussed in the text.}
	\label{fig:v2494cyg}
\end{figure*}

\subsection{IRAS 20566+5217 (V2494 Cyg) region}

Figure\,\ref{fig:v2494cyg} presents the immediate surroundings of IRAS\,20568+5217 (HH 381 IRS), which was suggested as the driving source of several Herbig-Haro (HH) objects \citep{devine:1997}. IRAS\,20568+5217 is also associated with a conical reflection nebula and has been shown to host an FU\,Ori type eruptive variable star \citep{aspin:2009}. This lead to the \object{V2494 Cyg} designation \citep{kazarovets:2011}. Our near-infrared colour-colour analysis (Fig.\,\ref{fig:ccdjhhk}) confirms the very young stellar nature of V2494\,Cyg.

An optical bipolar outflow (HH 965) has also been reported about 3 arcminutes to the south-west of V2494\,Cyg \citep{magakian:2010}. An embedded star found midway between the bow shocks that comprise HH\,965 has been proposed as the driving source; the IRAS HiRes-processed data reveals a 25$\mu$m peak at the position of this source (Fig.\,\ref{fig:v2494cyg}\,right panel), which is referred to as Z20566+5213 in the IRAS faint source reject catalogue \citep{moshir:2008}. The near-infrared colours of this object (Fig.\,\ref{fig:ccdjhhk}) are consistent with an embedded CTTS, perhaps even a low-mass pre-main-sequence (PMS) star.  This would explain the non-detection of the source in the mid-infrared.  We identify Z20566+5213 as the driving source of the HH\,965 outflow. 

There are several other HH knots and bow shocks observed in the close proximity of V2494\,Cyg (Fig.\,\ref{fig:v2494cyg}\,left panel). While HH\,381C belongs to the MHO\,900\,-\,904 north-south bipolar outflow (Table\,\ref{tab:outflows}), HH\,381A and B are likely being driven by a source situated in the north-east \citep{magakian:2010}.

\subsection{HH 380A region}

\begin{figure*}
	\centering
		\includegraphics[width=17cm]{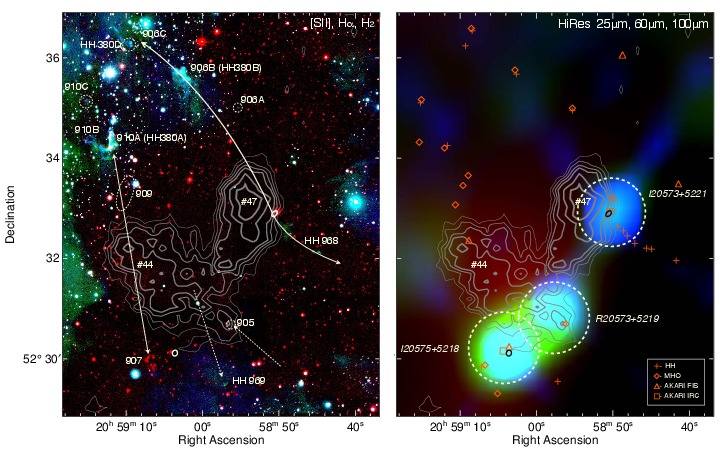}
	\caption{Similar to Fig.\,\ref{fig:v2494cyg} but for the area that includes HH\,380A (MHO\,910A).}
	\label{fig:hh380region}
\end{figure*}

Figure\,\ref{fig:hh380region} shows the region around HH\,380A and B, which were previously thought to belong to a single NE-SW outflow originating from V2494\,Cyg (known at that time as HH\,381~IRS) \citep{devine:1997}. Recently, however, \citet{magakian:2010} detected additional HH objects which support the existence of a number of outflows and therefore multiple driving sources (also theorised by \citet{devine:1997}). Our near-infrared data reveal numerous additional outflow features which, in conjunction with our multi-waveband data analysis, leads to a clear picture of the situation in this Figure.

\object{IRAS 20573+5221} appears to be the driving source of a bipolar outflow evident as MHO\,906 on one side and HH\,968 on the other (Fig.\,\ref{fig:hh380region}).  The HH\,968 lobe may extend as far as HH\,381A and B (shown in Fig.\,\ref{fig:v2494cyg}).  \object{IRAS 20573+5221} is identified as an embedded CTTS in our colour-colour analysis (Fig.\,\ref{fig:ccdjhhk}); it also coincides with the CSO 1.1mm core \#47 \citep{aspin:2011}.  This core is part of an extended dust continuum emission structure that connects with core \#44.  Both cores are only evident in far-infrared and millimetre wave data (Appendix\,\ref{sec:multisource}). Core \#44 coincides with an extended IRAS 100$\mu$m emission seen in the background (as red on Fig.\,\ref{fig:hh380region}\,right panel) and appears to be driving a 1.4pc-long bipolar outflow comprising MHO\,907, 909 and 910A. By contrast, the IRAS\,25 and 60$\mu$m emission peaks, labelled in Fig.\,\ref{fig:hh380region} as I20575+5218 and R20573+5219\footnote{From the IRAS point source reject catalogue - \citet{infrared-astronomical-satellite-iras-catalogs:2007}}, are situated on the outskirts of core \#44. This positional discrepancy is most likely the cause of the R20573+5219 source rejection. Careful inspection suggests that the source R20573+5219 may be connected with the yet unidentified 1.1mm sub-core (not labeled) in \#44 extended structure which can be clearly seen on the Fig.\,\ref{fig:hh380region}.

The region also contains several outflow features which also deserve mentioning. HH\,969 resembles a bow-shock which appears to be driven from a source associated with core \#44 or one of the IRAS peaks (I20575+5218 or R20573+5219). MHO\,905 (cf. Appendix\,\ref{sec:mhodetails}) is most likely the terminating bow of an outflow driven from the SW. Finally, MHO\,910B and 910C are aligned with HH\,968 and therefore may be associated with IRAS\,20573+5221. This could be explained with a strong precession and/or possible existence of binary component and therefore would support the original assumption of NE-SW outflow by \citet{devine:1997}.

\subsection{CN4 and CN5 region}

\begin{figure*}
	\centering
		\includegraphics[width=17cm]{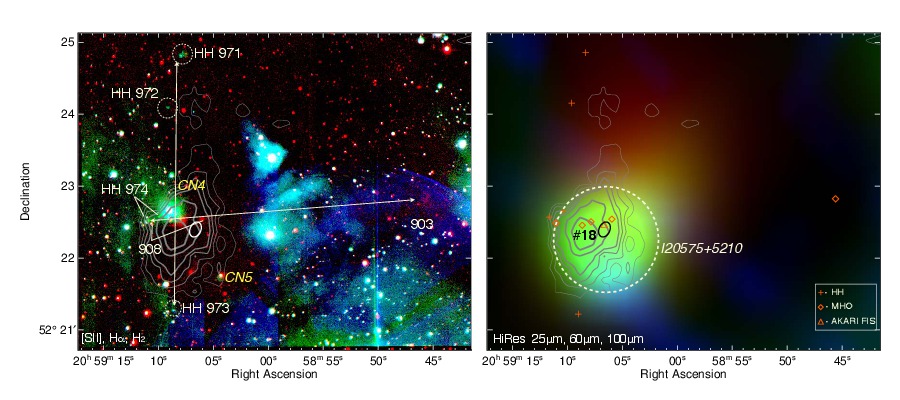}
	\caption{Similar to Fig.\,\ref{fig:v2494cyg} but for the area covering the CN4 and CN5.}
	\label{fig:cn4and5region}
\end{figure*}

Figure\,\ref{fig:cn4and5region} shows the area that includes the CN4 and CN5 cometary nebulae, as well as several HH objects discovered recently by \citet{magakian:2010}. Both nebulae have been suggested as the driving sources of the HH objects. Their near-infrared colours suggest that they are mildly embedded CTTS objects (Fig.\,\ref{fig:ccdjhhk}), although further spectroscopic study is needed further clarify their true nature. Based on the near-infrared H$_2$ 1-0 S(1) line observations and multicolour photometry presented in this paper, we are able to suggest an alternative scenario.

The area is host to a prominent IRAS source, \object{IRAS 20575+5210} (Fig.\,\ref{fig:cn4and5region}\,right panel).  IRAS\,20575+5210 also coincides with the CSO 1.1mm core \#18 \citep{aspin:2011}, which encompasses both CN4 and CN5. \citet{magakian:2010} suggest that CN4 is the driving source of \object{HH 974A} and \object{HH 974B}, but the most likely candidate appears to be IRAS\,20575+5210, which is only detected at far-infrared and millimetre wavelengths (Appendix\,\ref{sec:multisource}). In fact, this source appears to be driving a bipolar outflow consisting of MHO\,903, 908 and HH\,974 (Fig.\,\ref{fig:cn4and5region}). 

The same can be said about CN5 which was originally thought to be the driving source of \object{HH 973} \citep{magakian:2010}. However, our analysis suggests that HH\,973, together with \object{HH 972} and \object{HH 971}, comprise yet another bipolar outflow (Fig.\,\ref{fig:cn4and5region}, Table\,\ref{fig:outflows}) that is driven by IRAS\,20575+5210. It is likely that we are dealing with a multiple protostellar system embedded in a cold dust cloud identified as IRAS\,20575+5210.

\subsection{MHO 911}

\begin{figure*}
	\centering
		\includegraphics[width=18cm]{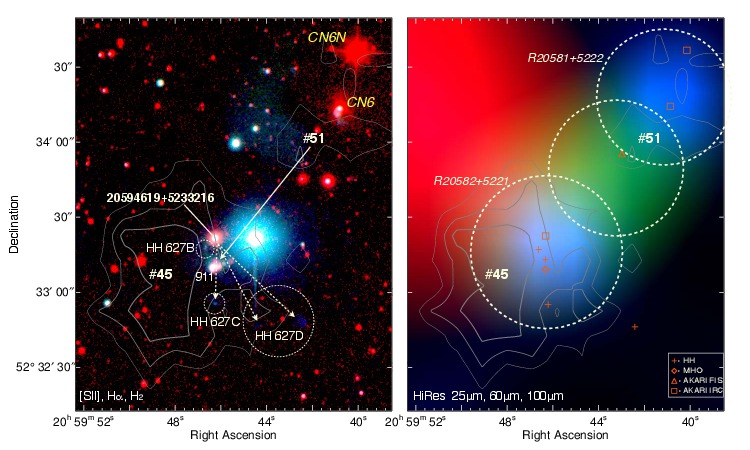}
	\caption{Similar to Fig.\,\ref{fig:v2494cyg} but for the area around MHO\,911 (HH\,627).}
	\label{fig:hh627region}
\end{figure*}

Figure\,\ref{fig:hh627region} shows the region around MHO\,911 and the optical knots that comprise \object{HH 627}, which were first detected by \citet{movsessian:2003}. The overall bow-shock shape of MHO\,911 suggests that it is driven by a source situated to the NW, where the CSO 1.1mm core \#51 is located \citep{aspin:2011}. Core \#51 appears to be associated with the 60$\mu$m IRAS emission feature R20581+5222; core \#51 also peaks 10-20\arcsec\ to the SE of the cometary nebula CN6. CN6 was reported to host an embedded CTTS \citep{aspin:2009}; our colour-colour analysis confirms this interpretation (Fig.\,\ref{fig:ccdjhhk}).  We therefore identify CN6 as the driving source of MHO\,911 (HH\,627) \citep{magakian:2010}. CN6N is not thought to be associated with MHO\,911, nor with R20581+5222: our colour analysis suggests that it is a background source, possibly an early-M type giant star  (Fig.\,\ref{fig:ccdjhhk}).

Identifying driving sources for the other shock features in Figure\,\ref{fig:hh627region} is more difficult.  IRAS\,R20582+5221 coincides with the 2MASS source 20594619+5233216, which has the colours of an embedded CTTS (it is labelled R20582+5221 on Fig.\,\ref{fig:ccdjhhk}).  This object could therefore be associated with some of the HH knots. Based on close inspection of the HH morphologies, we suggest that at least HH\,627C and D are driven by R20582+5221. The position of HH\,627B suggests that it could also be driven by R20582+5221, although CN6 is also a candidate if HH\,627B and MHO\,911 (HH\,627A) are parts of the same shock front.

\subsection{MHO 913}

\begin{figure*}
	\centering
		\includegraphics[width=17cm]{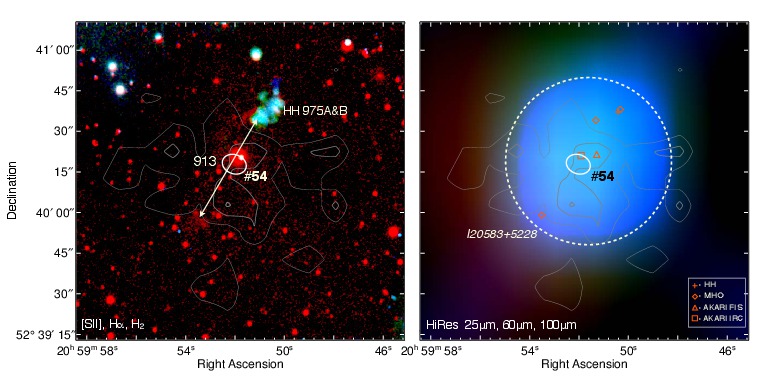}
	\caption{Similar to Fig.\,\ref{fig:v2494cyg} but for the MHO\,913 outflow.}
	\label{fig:mho913region}
\end{figure*}

Figure\,\ref{fig:mho913region} shows a new 0.2pc-long bipolar outflow originating from \object{IRAS 20583+5228}, which coincides with the CSO 1.1mm core \#54. The NW lobe of this bipolar outflow has been detected in the optical and labelled \object{HH 975} \citep{magakian:2010}, while the SW counter-lobe has only been observed in the near-infrared. Our colour-colour analysis confirms the YSO nature of \object{IRAS 20583+5228} (Fig.\,\ref{fig:ccdjhhk}), which is clearly the driving source of this outflow.

\subsection{Braid (V2495 Cyg) region}
\label{sec:braid}

\begin{figure*}
	\centering
		\includegraphics[width=17cm]{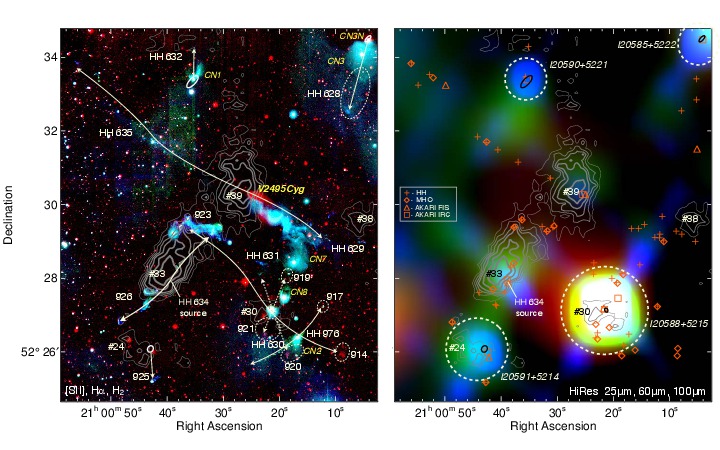}
	\caption{Similar to Fig.\,\ref{fig:v2494cyg} but for the area around the V2495\,Cyg (Braid Nebula).}
	\label{fig:braidregion}
\end{figure*}

The centre of our survey field (cf. Fig.\,\ref{fig:field-overview}) has been studied previously by  \citet{movsessian:2003}, although incomplete coverage in H$_2$ 1-0 S(1) emission limited their ability to identify outflows and their driving sources. Here, with the aid of wider-field near-IR observations and archival photometric data, we are more successful in understanding the situation.

Figure\,\ref{fig:braidregion} shows the immediate vicinity of the Braid Nebula overlaid with CSO 1.1mm emission contours \citep{aspin:2011}. The 1.1mm data clearly outline the densest regions (the cores) which most likely host the driving sources of the numerous outflows detected.

The Braid Nebula star, which coincides with core \#39 and a faint IRAS HiRes source, is identified as an FU Ori type outburst star \citep{movsessian:2006}.  It has recently received the designation \object{V2495 Cyg} \citep{kazarovets:2011}.  \object{V2495 Cyg} drives a bipolar outflow that is traced by \object{HH 635} in the NE and \object{HH 629} in the SW. The apparent asymmetry of the bipolar outflow is most likely caused by the presence of dense, cold dust to the NE (Fig.\,\ref{fig:braidregion}).

As can be seen from Fig.\,\ref{fig:braidregion}, one of the most prominent sources in the region is \object{IRAS 20588+5215} (IRAS\,15), which coincides with the CSO 1.1mm core \#30. Since the IRAS emission associated with this source is quite extended, it encompasses several sources which we will discuss separately for the sake of clarity. Optical and near-infrared observation reveal two stars (IRAS 15N and IRAS 15S) in the vicinity of the IRAS error ellipse. IRAS\,15N was spectroscopically studied by \citet{aspin:2009} who concluded that it is a highly active accreting CTTS with a significant K-band thermal excess. Our analysis suggest that IRAS 15N is driving at least one bipolar outflow spanning from MHO\,914 to 923 (\object{HH 448}, \object{RNO 127}). IRAS\,15N might be responsible for two other bipolar outflows, traced by MHO\,920 to \object{HH 631} and by MHO\,921 to 919, as shown in Fig.\,\ref{fig:braidregion}. IRAS\,15S, by contrast, was found to be a foreground object.  A near-IR spectrum obtained toward IRAS\,15S does contain H$_2$ emission \citep{aspin:2009}; we suggest that this could be caused by the superimposition of outflow features emanated from IRAS\,15N.

The CN2 and CN8 cometary nebulae are situated within the extent of IRAS\,15 (cf. Fig.\,\ref{fig:braidregion}). Our colour-colour analysis indicates that both are CTTSs (Fig.\,\ref{fig:ccdjhhk}). \citet{magakian:2010} suggest that CN2 drives the outflow traced by MHO\,917 and \object{HH 630}. This seems likely given the positional placement of the detected outflow features. There is no mid-infrared emission detected towards CN2, though this could be due to the AKARI/IRC detection limits (50mJy $\sim$ 7.6 mag at 9$\mu$m). Near-IR spectroscopic data presented by \citet{aspin:2009} suggests that at one stage CN8 drew wide-angle molecular outflow \citep{magakian:2010} suggesting a different scenario for the MHO\,919 origin.

The area presented in Fig.\,\ref{fig:braidregion} is also host to a prominent optical and near-infrared flow, MHO\,926 (\object{HH 634}). MHO\,926 consists of a chain of well-defined bows connecting with MHO\,923A and B \citep{movsessian:2003}. It seems clear that the bipolar outflow MHO\,926 to 923A,B is being powered by a source associated with the 1.1mm core \#33.  This driving source is detected in the AKARI/IRC and AKARI/FIS catalogues (Appendix\,\ref{sec:multisource}).  Moreover, the near-IR colours of this object (J - H = 6.09 and H - K = 5.95;  not shown on Fig.\,\ref{fig:ccdjhhk}) indicate that it is deeply embedded. It is interesting to note that the MHO\,923/926 outflow appears to terminate in the outskirts of the core that hosts V2495\,Cyg (core \#39).  This could simply be due to increased extinction as the flow drills into or passes behind core \#39.

There are several other monopolar flows in the area which deserve mentioned. \object{IRAS 20591+5214} (IRAS 14) drives a short flow represented by MHO\,925 (\object{HH 633}) \citep{movsessian:2003}. The source itself has the typical colours of a Class\,II YSO or possibly a Class\,I protostar.  

\object{IRAS 20590+5221} or the associated cometary nebula CN1 was classified as an accreting CTTS star \citep{aspin:2009} which drives \object{HH 632} \citep{magakian:2010}. Unfortunately we did not detect any near-infrared H$_2$ 1-0 S(1) line features associated with this outflow, although our colour analysis does confirm the YSO nature of CN1 (Fig.\,\ref{fig:ccdjhhk}). 

Finally, the monopolar flow, \object{HH 628}, is driven by an embedded CTTS object, CN3N (\object{IRAS 20585+5222}).

\subsection{Cyg 19 region}

\begin{figure*}
	\centering
		\includegraphics[width=17cm]{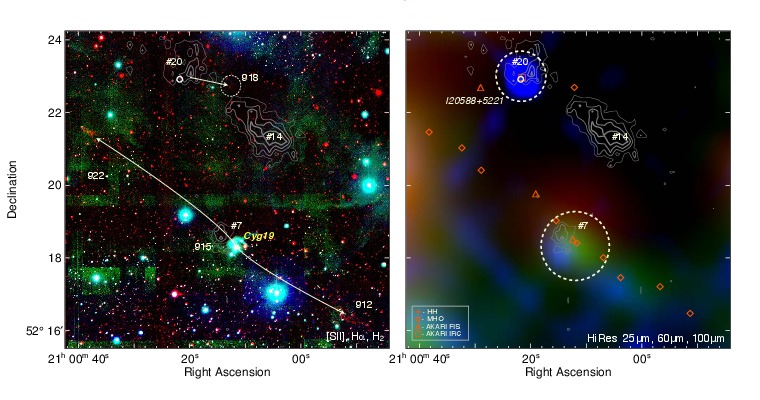}
	\caption{Similar to Fig.\,\ref{fig:v2494cyg} but for the Cyg\,19 region.}
	\label{fig:cyg19region}
\end{figure*}

Figure\,\ref{fig:cyg19region} shows the area around Cyg 19, an H$\alpha$ star first reported by \citet{melikian:1996}. Our near-infrared H$_2$ line survey revealed a 2.3pc-long (Table\,\ref{tab:outflows}) bipolar outflow comprising MHO\,912, 915 and 922.  Each MHO represents a chain of bows and arcs. Cyg\,19 is situated on the axis that defines the lobe and counter-lobe of this bipolar outflow. It could be the driving source of this flow, although its near-infrared colours and spectroscopy  \citep{aspin:2009} suggest that it is a mildly embedded dwarf star. The presence of curving nebulosity around Cyg\,19 indicates an outflow cavity seen close to pole-on that is illuminated by scattered light. This nebulosity, together with the detection of H$\alpha$ emission \citep{melikian:1996}, indicates that Cyg\,19 could be a sporadically accreting YSO \citep{aspin:2009}.  Although the CSO 1.1mm core \#7 does not coincide exactly with Cyg\,19, there is faint extended IRAS emission towards this source. The former could be due to the positional uncertainty of the CSO 1.1mm data \citep{aspin:2011}, or the 1.1mm data could indicate the true location of the MHO\,912/915/922 bipolar outflow driving source.

About 4 arc-minutes north we identify a faint H$_2$ feature, MHO\,918 (Fig.\,\ref{fig:cyg19region}), which could be driven from either CSO 1.1mm core \#14 or core \#20 \citep{aspin:2011}.  Core \#20 is perhaps the most likely source, since it coincides with \object{IRAS 20588+5221}. There is also a star in the centre of core \#20 detected in the near- and mid-infrared with colours typical of a heavily embedded protostar (possibly a Class\,I source; Fig.\,\ref{fig:ccdjhhk}).

\subsection{IRAS 21005+5217 region}

\begin{figure*}
	\centering
		\includegraphics[width=17cm]{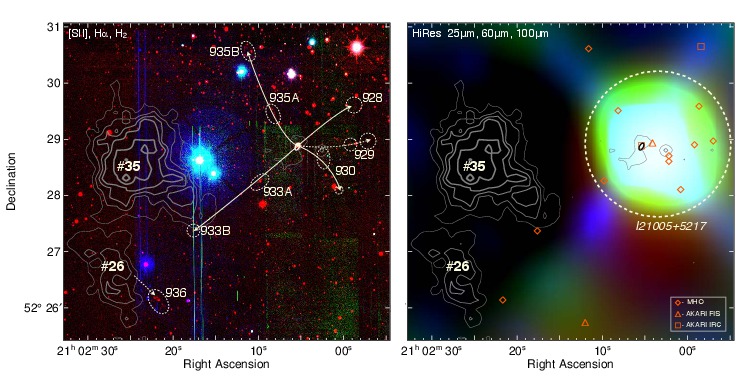}
	\caption{Similar to Fig.\,\ref{fig:v2494cyg} but for the IRAS\,21005+5217 region.}
	\label{fig:mho928region}
\end{figure*}

Figure\,\ref{fig:mho928region} illustrates the area containing MHO\,928, 929, 930, 933 and 936. Our analysis suggest that there are two bipolar, almost perpendicular outflows emanating from \object{IRAS 21005+5217}.  One spans from MHO\,930 to 935 while the other includes MHO\,933 and 928. It is possible that the NW side of the second flow also contains MHO\,929 if this outflow is precessing (Table\,\ref{tab:outflows}). Our near-infrared colour-colour analysis (Fig.\,\ref{fig:ccdjhhk}) suggests that \object{IRAS 21005+5217} is an embedded CTTS or possibly a younger object.  This source is prominent in the IRAS HiRes 100$\mu$m data (Fig.\,\ref{fig:mho928region}). However, there is no CSO 1.1mm core associated with IRAS 21005+5217  \citep[within the 3$\sigma$ detection limit][]{aspin:2011}.  This may be explained by an absence of dust and gas, due to the presence of a high energy intermediate or high mass source.

There are two other 1.1mm cores in Fig.\,\ref{fig:mho928region}, which could be the driving sources of MHO\,936. The location and the overall morphology of MHO\,936 suggests that it has been driven from core \#26.

\subsection{MHO 931 and 932}

\begin{figure*}
	\centering
		\includegraphics[width=17cm]{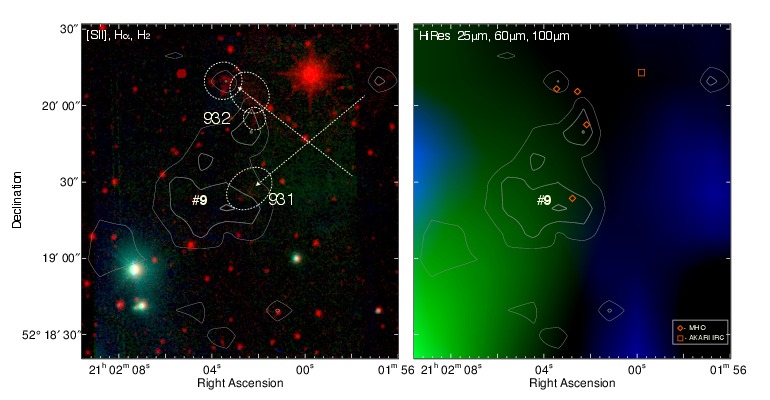}
	\caption{Similar to Fig.\,\ref{fig:v2494cyg} but for the area around MHO 932 and 931.}
	\label{fig:mho932region}
\end{figure*}

Figure\,\ref{fig:mho932region} shows the region where MHO\,931 and 932 are located. The morphology of these objects suggests different outflow directions as indicated by the arrows. MHO\,931 might originate from a source to the north-west, while MHO\,932 might be driven from the south-west. The bright star 2MASS\,21015977+5220127 evident in this figure has no apparent morphological feature which might suggest that it is the driving source of the objects. Moreover, our colour-colour analysis suggests that it is a background source (Fig.\,\ref{fig:ccdjhhk}). The presence of these two MHOs -- 931 and 932 -- in close proximity could be explained by their encounter with the dense environment associated with CSO 1.1mm core \#9.   

\subsection{MHO 934}

\begin{figure*}
	\centering
		\includegraphics[width=8.5cm]{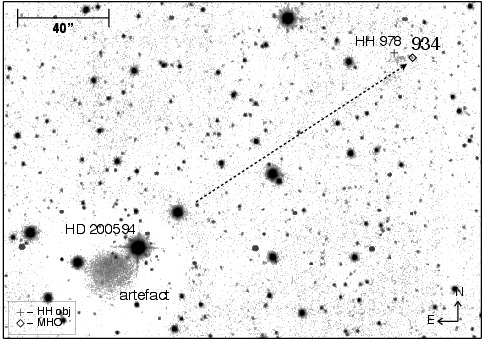}
	\caption{Near-infrared H$_2$ 1-0 S(1)+Continuum view of the area around MHO 934. Dashed-line arrow shows the suspected direction of the flow.}
	\label{fig:mho934}
\end{figure*}

MHO\,934 (Fig.\,\ref{fig:mho934}) is situated in an area of our surveyed field where we lack complete coverage in all optical bands (Fig.\,\ref{fig:field-overview}).  Never-the-less, it is associated with \object{HH 978} \citep{magakian:2010}. The overall morphology of this bow shock suggests a driving source to the SE, where \object{HD 200594} is located. However, the near-infrared colours of this source indicate that it is a main sequence star (Fig.\,\ref{fig:ccdjhhk}). It is therefore unlikely to be the driving source of the MHO\,934 bow shock.
	
\end{appendix}

\begin{appendix}
\onecolumn

\section{Details of the detected MHOs}
\label{sec:mhodetails}
\begin{longtable}{p{101mm}p{60mm}}
	\caption{Detailed descriptions of the sub-regions where MHOs (see Tab.\,\ref{tab:mhos}) were identified. In each case images are presented in H$_2$ 1-0 S(1) line emission before (left) and after (right) continuum subtraction. The positions of known Herbig-Haro (HH) objects, Cometary Nebulae (CN) and IRAS sources are indicated.\label{tab:mhodet}}\\
	\hline\hline
	Images &  Description\\
	\hline
	\endfirsthead
	\caption{continued.}\\
	\hline\hline
	Images &  Description\\
	\hline
	\endhead
	\hline
	\endfoot
\newline \includegraphics[width=10cm]{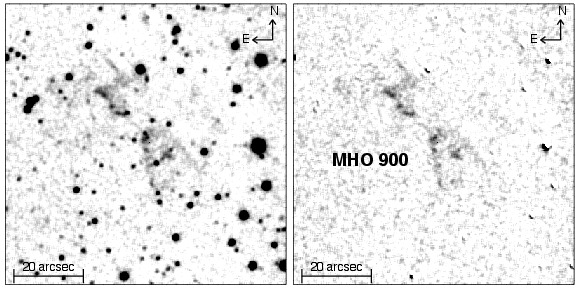} & 
\newline Area including MHO\,900. Despite the overall resemblance to an edge-on disk system with a bipolar outflow cavity we were unable to identify any IRAS or AKARI sources here. The system is most likely a broken bow-shock driven from V2494\,Cyg.\\
\hline
\newline \includegraphics[width=10cm]{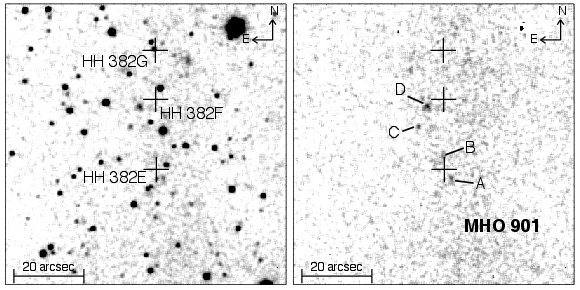} & 
\newline Area including the MHO\,901 complex coinciding with the optical HH\,382G, F and E knots.\\
\hline
\newline \includegraphics[width=10cm]{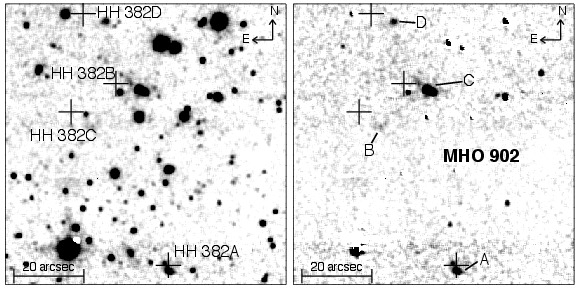} & 
\newline Area including the MHO\,902 knots which coincide with the optical HH\,382A, B and C knots.\\
\hline
\newline \includegraphics[width=10cm]{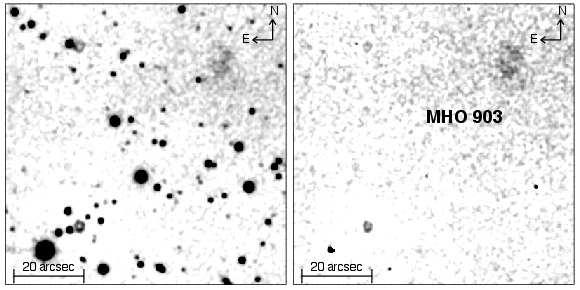} & 
\newline This sub field includes MHO\,903 which is rather diffuse in structure.\\
\hline
\newline \includegraphics[width=10cm]{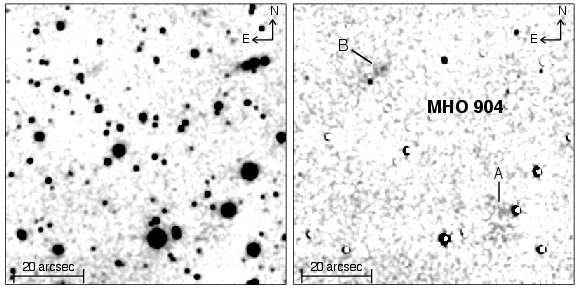} & 
\newline The area shows the MHO\,904 knots which belong to the outflow emanated from V2494\,Cyg. These are the most southern knots discovered in our survey.\\
\hline
\newline \includegraphics[width=10cm]{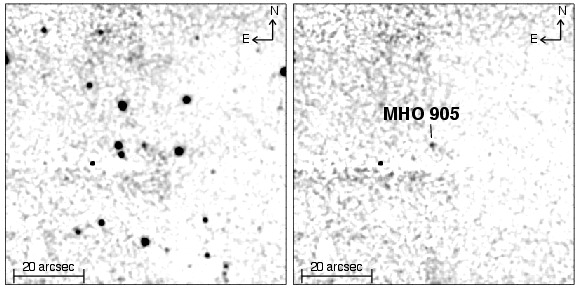} & 
\newline MHO\,905 is a small bow-shock, the origin of which is still unknown.\\
\hline
\newline \includegraphics[width=10cm]{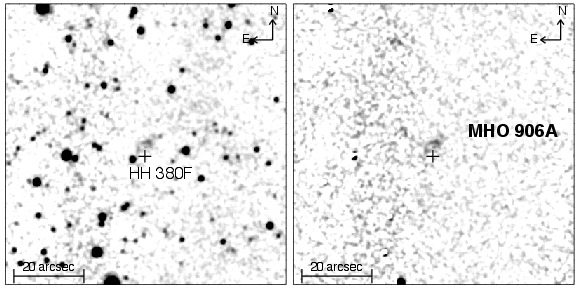} & 
\newline This is one of the knots in the MHO\,906 outflow.  This feature is associated with HH\,380F.\\
\hline
\newline \includegraphics[width=10cm]{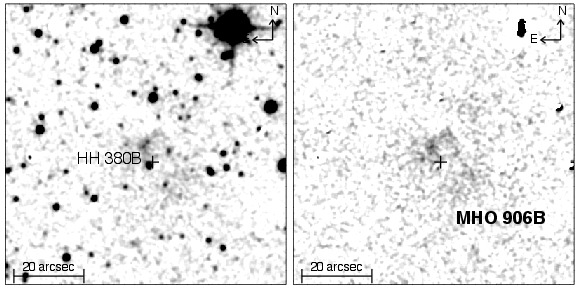} & 
\newline Part of the MHO\,906 outflow coinciding with HH\,380B.\\
\hline
\newline \includegraphics[width=10cm]{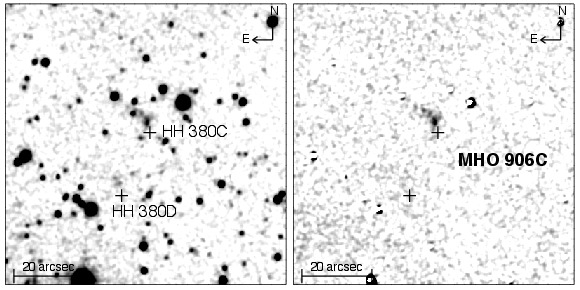} & 
\newline Last part of the MHO\,906 outflow coinciding with the optical feature HH\,380C. Morphologically it resembles a bow-shock driven from the South-East. 
In this current work we assume an outflow direction to the  South-West.\\
\hline
\newline \includegraphics[width=10cm]{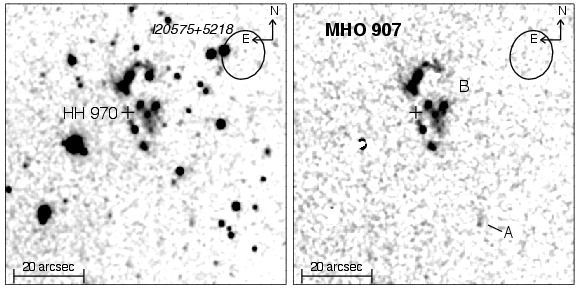} & 
\newline Area includes the MHO\,907 complex bow shock which coincides with the faint object HH\,970. IRAS\,20575+5218 is unrelated to this flow.\\
\hline
\newline \includegraphics[width=10cm]{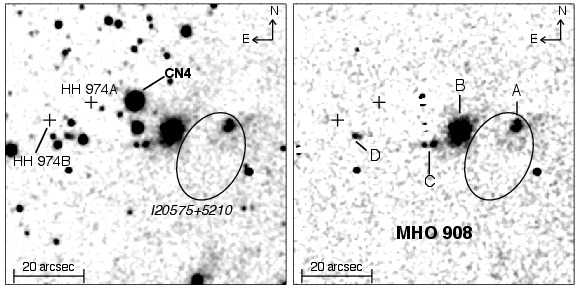} & 
\newline Region including the MHO\,908 chain of knots A to D, where knot D might be related to HH\,974B. The IRAS\,20575+5210 error ellipse and CN4 are marked for reference. \\
\hline
\newline \includegraphics[width=10cm]{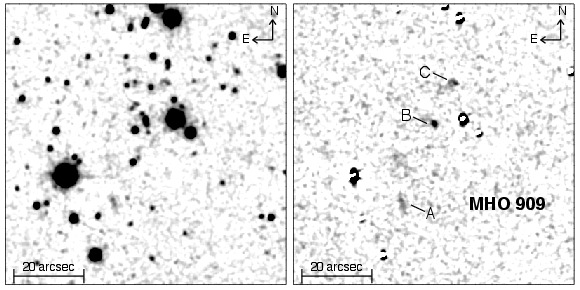} & 
\newline Area shows a chain of relatively faint knots comprising MHO\,909.\\
\hline
\newline \includegraphics[width=10cm]{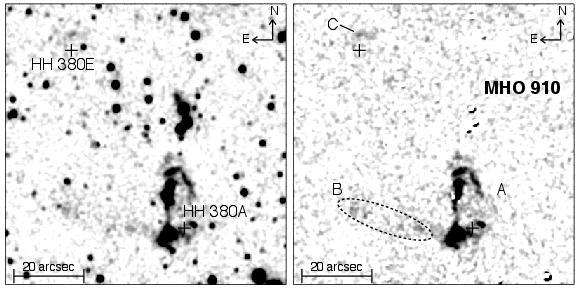} & 
\newline Area shows the details of MHO\,910, which is associated with the prominent optical bow-shock HH\,380(A). There is a faint extension to the East-North-East (MHO\,910B) which could be unrelated to this flow. MHO\,910C also coincides with HH\,380E.\\
\hline
\newline \includegraphics[width=10cm]{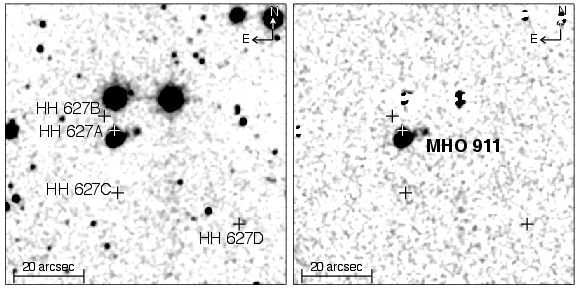} & 
\newline Region where the MHO\,911 bow-shock can be seen. This quite prominent and compact bow also coincides with HH\,627A.\\
\hline
\newline \includegraphics[width=10cm]{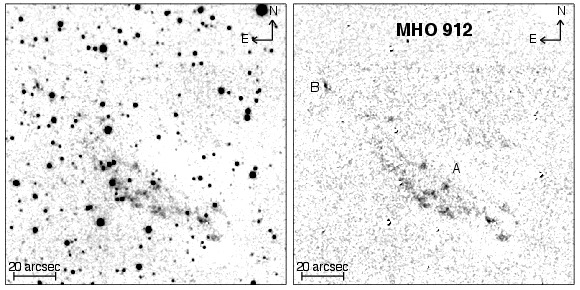} & 
\newline Figure shows the complex bow-shock structure MHO\,912, which is part of a prominent bipolar outflow that is only detected in the near-infrared.\\
\hline
\newline \includegraphics[width=10cm]{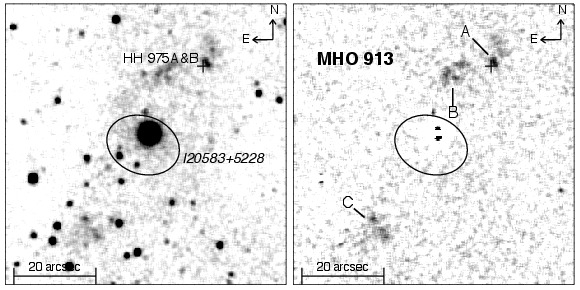} & 
\newline Area including the compact bipolar outflow MHO\,913, emanating from IRAS\,20583+5228. Knots A and B correspond to HH\,975A and B.\\
\hline
\newline \includegraphics[width=10cm]{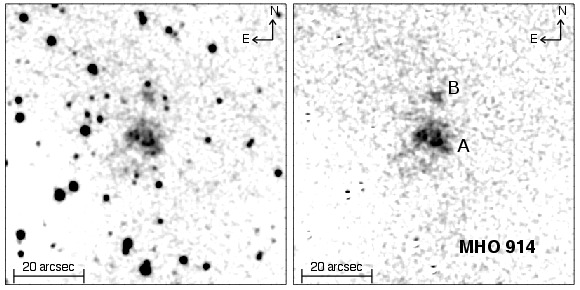} & 
\newline Figure shows the complex bow-shock structure of MHO\,914.\\
\hline
\newline \includegraphics[width=10cm]{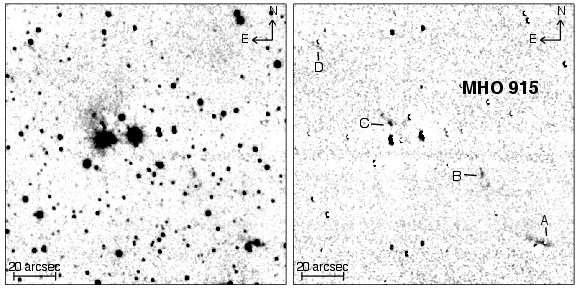} & 
\newline Area including the MHO\,915 chain of knots emanating from Cyg\,19.  These are part of a prominent near-infrared bipolar outflow.\\
\hline
\newline \includegraphics[width=10cm]{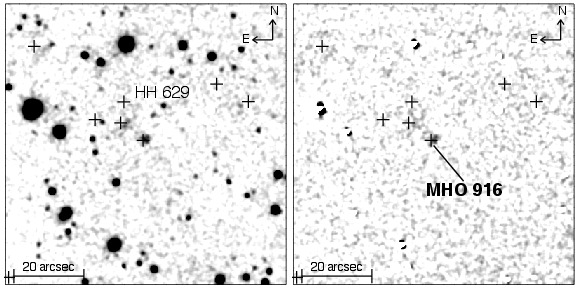} & 
\newline Region including the faint knot MHO\,916 which coincides with HH\,629 and emanates from V2495\,Cyg.\\
\hline
\newline \includegraphics[width=10cm]{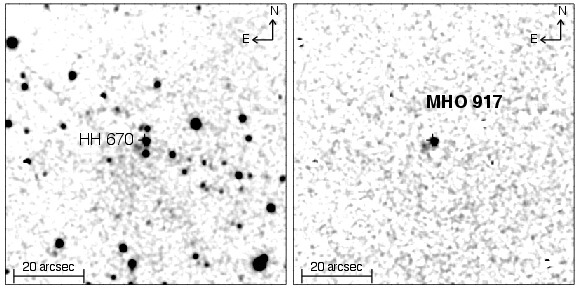} & 
\newline The compact knot MHO\,917 coinciding with HH\,670.\\
\hline
\newline \includegraphics[width=10cm]{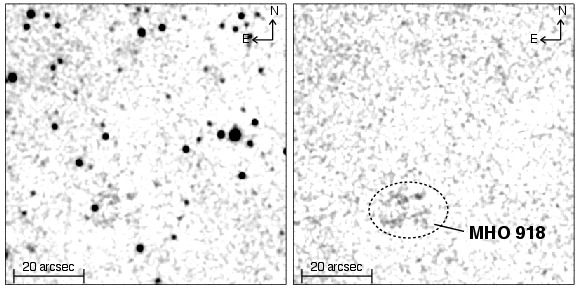} & 
\newline The faint diffuse bow-shock MHO\,918.\\
\hline
\newline \includegraphics[width=10cm]{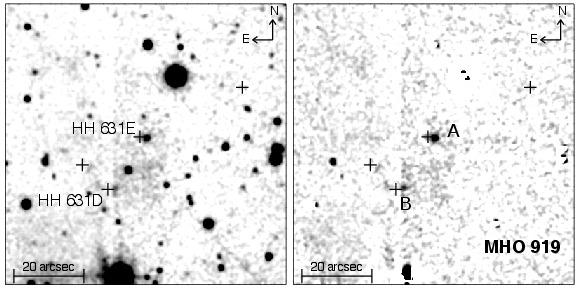} & 
\newline MHO\,919 knots also coinciding with the HH\,631 knots D and E.\\
\hline
\newline \includegraphics[width=10cm]{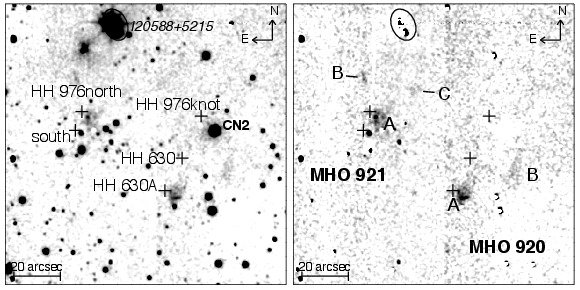} & 
\newline Area including MHO\,920 and 921 which is also full of HH objects, some coinciding with the near-infrared features. 
The IRAS\,20588+5215 (IRAS\,15) error ellipse is shown, as is the position of CN2.\\
\hline
\newline \includegraphics[width=10cm]{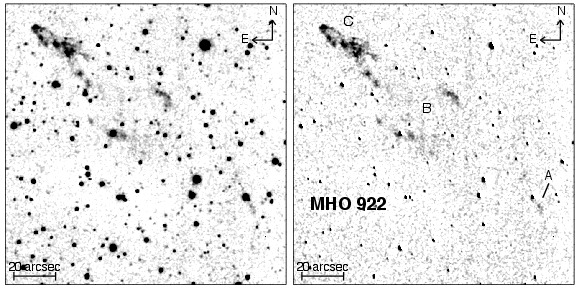} & 
\newline Area including the extensive and rather complex chain of bow-shocks MHO\,922, which is part of a near-infrared bipolar outflow.\\
\hline
\newline \includegraphics[width=10cm]{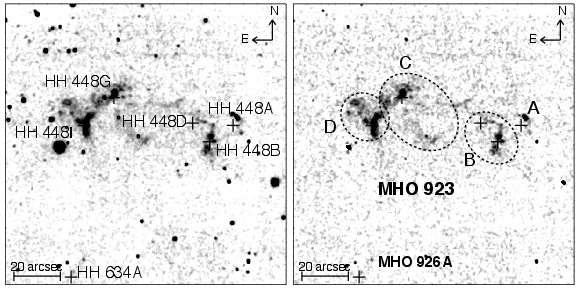} & 
\newline Region showing the complex object MHO\,923, which coincides with HH\,448 and is also known as RNO\,127. MHO\,926A, which coincides with HH\,634A, can also be seen in this image.\\
\hline
\newline \includegraphics[width=10cm]{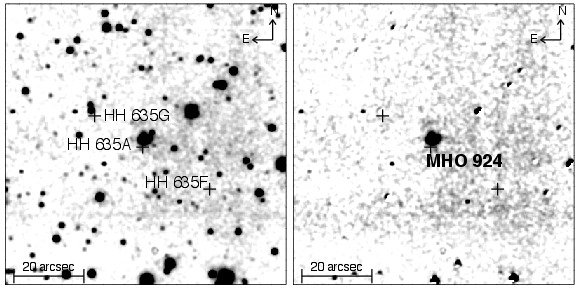} & 
\newline Area shows the MHO\,924 bright knot which coincides with HH\,635A.\\
\hline
\newline \includegraphics[width=10cm]{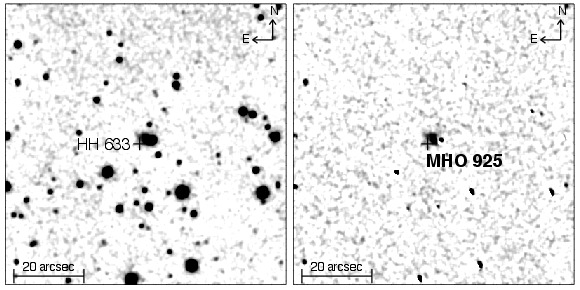} & 
\newline Area shows MHO\,925 which coincides with HH\,633.\\
\hline
\newline \includegraphics[width=10cm]{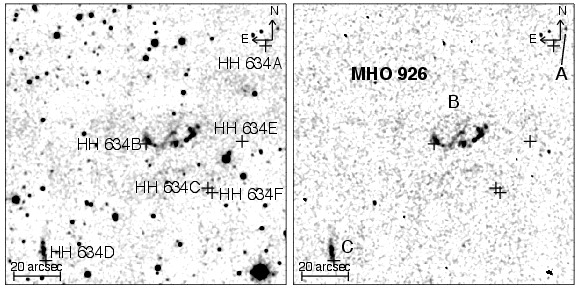} & 
\newline Area shows the MHO\,926 outflow, which also coincides with the optical HH\,634 outflow.\\
\hline
\newline \includegraphics[width=10cm]{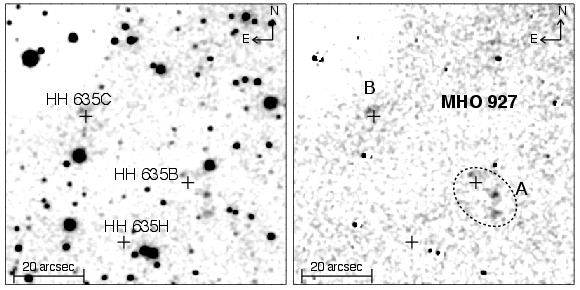} & 
\newline Region shows the MHO\,927 group of knots and bow-shocks which coincide with HH\,635B and C. This is part of the flow driven by V2495\,Cyg.\\
\hline
\newline \includegraphics[width=10cm]{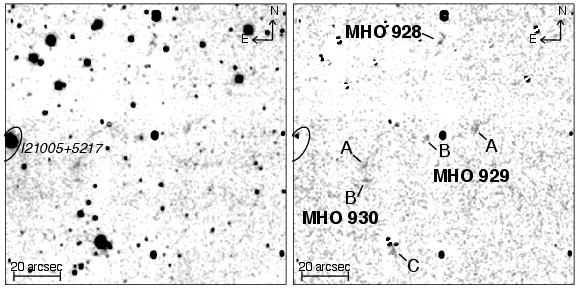} & 
\newline Group of bows and knots emanating from IRAS\,21005+5217. MHO\,928 and 929 are thought to be the part of a West-to-East or East-to-South bipolar flow. MHO\,930 is part of a bipolar outflow extending in a SW-NE direction.\\
\hline
\newline \includegraphics[width=10cm]{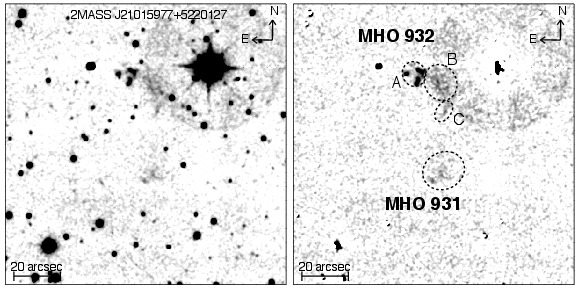} & 
\newline Area including MHO\,932 and 931. These two objects are thought to belong to different flows.\\
\hline
\newline \includegraphics[width=10cm]{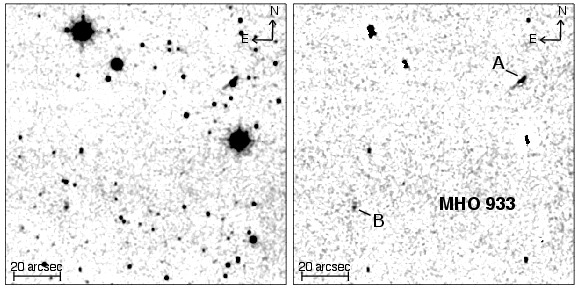} & 
\newline Region shows the MHO\,933 knots which are thought to be driven by IRAS\,21005+5217.  MHO\,933 belongs to the same bipolar outflow as MHO\,928 and 929.\\
\hline
\newline \includegraphics[width=10cm]{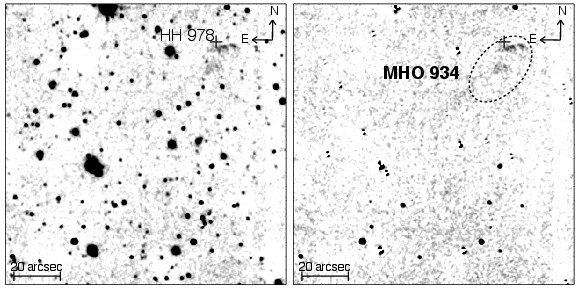} & 
\newline The area shows the MHO\,934 chain of bow-shocks which are thought to be driven by a source somewhere in the SE.\\
\hline
\newline \includegraphics[width=10cm]{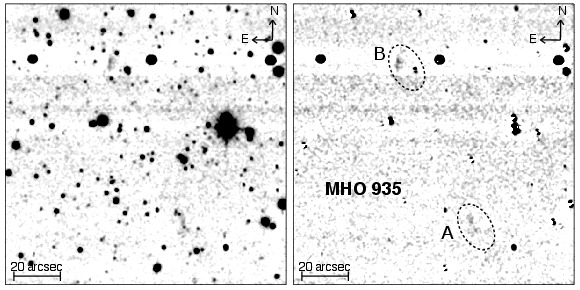} & 
\newline Region shows MHO\,935 which is driven by IRAS\,21005+5217. MHO\,935   belongs to the same bipolar outflow as MHO\,930.\\
\hline
\newline \includegraphics[width=10cm]{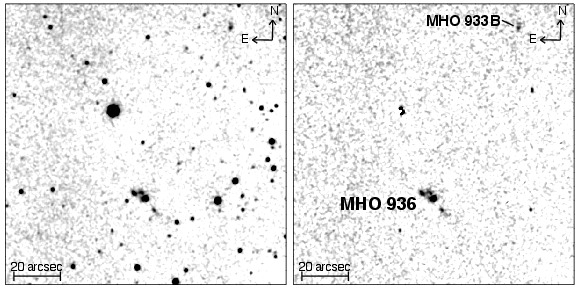} & 
\newline Area shows MHO\,936 which has quite a complex structure.  It is thought to be driven by a source somewhere in the North-East.\\
\end{longtable}

\end{appendix}


\end{document}